\documentclass[preprint]{aastex}

\newcommand{\mgal}{\ensuremath{\times 10^{11}M_{\odot}}}

\newcommand{\fig}{Figure }
\newcommand{\cmr}{\ensuremath{\mathbf{r}}}
\newcommand{\Gam}{\ensuremath{\mathbf{\Gamma}}}
\newcommand{\Ps}{\ensuremath{\mathbf{\Psi}}}
\newcommand{\I}{\ensuremath{\mathbf{\I}}}
\newcommand{\ML}{\ensuremath{\mathcal{M/L}}}
\newcommand{\kpc}{\ensuremath{\mathrm{kpc}}}
\newcommand{\etag}{\ensuremath{\eta_{G}\ }}
\shorttitle{Theoretical Modeling of Polarized Sources}
\shortauthors{}

\begin{document}

\title{Theoretical Modeling of Weakly Lensed Polarized Radio Sources}

\author{Christopher R. Burns}
\affil{Swarthmore College, Swarthmore, PA 19086}
\email{cburns1@swarthmore.edu}

\author{Charles C. Dyer}
\affil{University of Toronto, Toronto Ontario Canada, M5S 3H8}
\email{dyer@astro.utoronto.ca}

\author{Philipp P. Kronberg}
\affil{Los Alamos National Laboratory, IGPP, MS C305, Los Alamos, NM 87501}
\email{kronberg@lanl.gov}

\and{}

\author{Hermann-Josef R\"{o}ser\altaffilmark{1}}
\affil{Max Planck Institut f\"{u}r Astronomie, Heidelberg, Germany}
\altaffiltext{1}{Visiting Astronomer, German-Spanish Astronomical Centre,
Calar Alto, operated by the Max-Planck-Institute for Astronomy, Heidelberg,
jointly with the Spanish National Commission for Astronomy.}
\email{roeser@mpia-hd.mpg.de}

\begin{abstract}
In this paper we present the theoretical basis for the modeling of
weakly gravitationally lensed extended sources that are polarized. This 
technique
has been used in the past to constrain the mass profiles of galaxies
projected against
the jet of the quasar 3C9. Since then, work has been done to improve
both the measurement and theoretical modeling of the lensing signal,
which manifests itself as an alignment breaking between the morphology
and the polarization, parametrized as \etag. To this end, we present 
the mathematical
derivation of the theoretical value of \etag as well as numerical
simulations of expected signals in polarized radio jets. We use 
the radio jet sources 3C9
and 1253+104 as illustrative examples of the measurement and modeling of the
\etag signal.  For 3C9, we present constraints on the parameters
of the two intervening lenses and quantify their confidence intervals.  One
lens has no measured redshift and in this case, we show the dependence of
mass and mass-to-light ratio on assumed redshift.

\end{abstract}

\keywords{gravitational lensing --- polarization ---
galaxies: fundamental parameters --- galaxies: jets --- 
quasars:  individual (3C9, QSO1253+104)}

\section{INTRODUCTION}

The action of a gravitational lens is to map the source plane to the
image plane. This mapping can either be one-to-many (multiple images)
or one-to-one (weak lensing). In the majority of cases, we do not
know the structure of the source plane and must make some assumptions
about it in order to invert the lens equation and solve for the lens
parameters. However, if the source is extended and polarized, then we can have 
an undistorted
view of the structure of the unlensed source plane since the polarization, being
a one-point vector,
is unchanged by a gravitational lens \citep{Kronberg:91,Dyer:92}.
The morphology, however,  is distorted due to the differential
mapping of different points in the source plane, which represent a family
of 2-point vectors. The fact that a source
is polarized indicates a preferred direction in the source plane which
is often
aligned with, or physically coupled to, the morphology.  
Astrophysical jets are believed to
originate from strong magnetic fields which accelerate charged particles
and produce the synchrotron radiation that makes the 
jet's morphology visible \citep{Honda:2002,Colgate:2001,Benford:78}.  It is 
therefore 
expected
that the polarization should be aligned with the jet direction and this is
observed to be the case \citep{Bridle-Perley:86}.

A key advantage of this method is that it provides the possibility of 
uniquely probing the global mass and 
mass distribution of a single galaxy, rather than by statistical methods 
that use collections of 
individual galaxy lenses. The polarimetric properties of radio sources 
provide unique advantages for this
(see below), one being that the linear scales of extragalactic radio source
jets are comparable with a single galaxy's mass scale. The method could
measure all the mass included within a galactocentric radius,
$r \lesssim $ 100 \kpc. 

A disadvantage is that the ``setup'' for this type of single-galaxy mass 
probe requires some serendipity in getting
advantageous lens-source alignments, examples of which we discuss in this paper. Up to now these are generally few, being limited 
by the sensitivity of the immediate past generation 
of large optical telescopes. For the ideal galaxy mass probe, quantitative light profiles of the lensing galaxy, with colors, 
on scales comparable with the background radio jet images would give 
estimates of the mass-to-light ratio (\ML). This sort of detail will become 
possible with the future NGST optical, and EVLA radio telescopes, and more powerful ground-based optical telescopes. 
It is therefore timely at this point to examine the lensing theory aspects 
for radio sources, which have 
yet to come to full fruition, but are potentially very powerful single-galaxy
mass probes.

A source can be characterized by two vector fields:  its morphology defined by
Stokes parameter I, and that by the linear polarization (Stokes Q and U).
The  morphological
features will be mapped as two-point vectors and will rotate.
Therefore, if the two vector fields were aligned
in the source plane, any differential gravitational lensing will cause
the breaking of this alignment by an angle we call \etag.  This has 
been measured and used to
constrain the mass of galaxy-scale lenses in front of the jet of quasar 3C9
\citep{Kronberg:91,Kronberg:96}. More generally, it could be used to measure 
the 
gravitational
lensing of any source where morphology and polarization are intrinsically
correlated in
the source plane.

In the following sections, we present the theoretical derivation of
the alignment breaking parameter \etag for the case of a linear
morphology applicable to astrophysical jets and the
peripheral boundaries of radio lobes. We follow with numerical 
simulations for
an ideal jet to illustrate the measurement and modeling of the alignment
breaking. We then apply these new techniques to the quasar 3C9 and derive 
masses and mass-to-light ratios for the two intervening galaxies. Lastly, we present the observations
and analysis
of the quasar 1253+104, which shows no alignment breaking and is
consistent with the observed redshift of the intervening galaxy.

\section{DERIVATION OF \etag}
\begin{figure}
\plotone{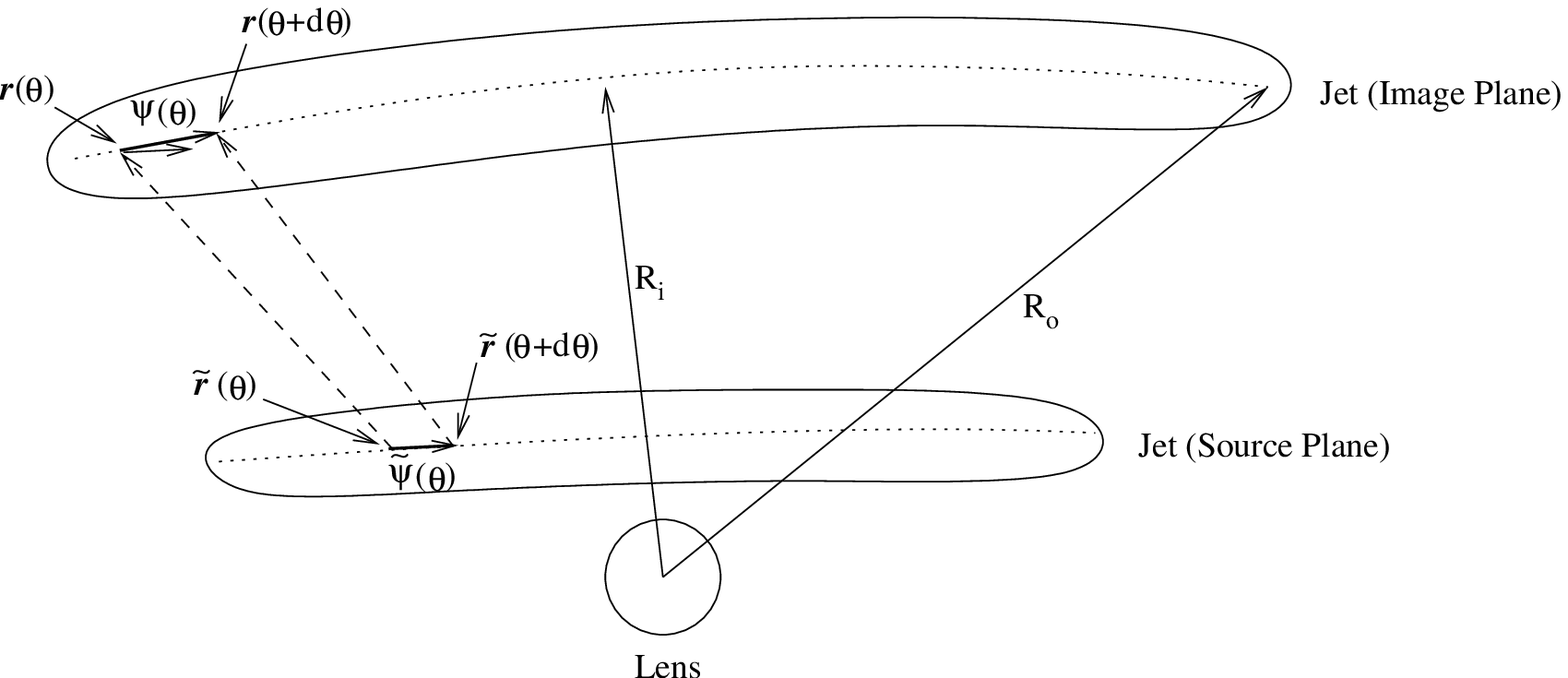}
\figcaption{The lensed (image plane) and unlensed (source plane) images of
a source with one dimensional morphology.  The dotted lines represent the
fiducial lines.  The dashed lines show the mapping between two points in
the source plane and their corresponding points in the image plane.  Note
that the tangent vector in the source plane ($\tilde{\mathbf{\Psi}}$) has
been reproduced below the tangent vector in the image plane ($\mathbf{\Psi}$)
to illustrate the alignment breaking.  The minimum and maximum impact parameters
probed by the jet ($R_i$ and $R_o$, respectively) are drawn with arrows.
\label{fig:etag_defined}}
\end{figure}

In order to measure the \etag signal, one must obtain representative vectors
which define the local morphology
of the source.  In the case of astrophysical jets, which are typically
one-dimensional (i.e., the transverse angular size of the jet is much less than
its linear extent or even the resolving power of the telescope), the tangent 
vector is used.  We therefore
ascribe a curve $\Gam \left(\theta \right)$, which
we call the fiducial curve, which best defines the path
of the jet or lobe boundary (see 
\fig \ref{fig:etag_defined}). At each
point, one can then measure the tangent vector 
$\Ps \left(\theta \right)=d\Gam /d\theta $
and the intrinsic polarization angle $\chi \left(\theta \right)$. For an ideal alignment
and no gravitational lensing, one expects 
the angle of the tangent vector to be aligned with the polarization 
($\psi \left(\theta \right)=\chi \left(\theta \right)$,
where $\psi$ is the angle of the tangent vector).
In practice, it is more likely there is some intrinsic deviation from
the ideal case and we therefore should write 
$\chi \left(\theta \right)=\psi \left(\theta \right)+\kappa \left(\theta \right)$,
where $\kappa $ represents any intrinsic deviation from perfect alignment
in the source plane. With these definitions in hand, the
alignment breaking parameter is defined \citep{Kronberg:91}:
\begin{equation}
\eta _{G}=\chi \left(\theta \right)-\psi \left(\theta \right)+\kappa 
\left(\theta \right)\label{Eq:etag}
\end{equation}
 such that in the ideal case with no gravitational lensing, we expect
$\eta _{G}=0$ over the length of the structure.

It is convenient to think of the projected positions of the lenses
and sources on the sky as complex numbers. Following the complex notation
of \citet{Bourassa_Kantowski:1973}, we label image-plane positions
as $\cmr$ and the source-plane positions as $\tilde{\cmr}$. The action
of a gravitational lens is therefore to map each source-plane point
to one or more points on the image plane. This map can be written
as
\begin{equation}
\tilde{\cmr}=\cmr-\sum _{l}D_{l}\: \mathbf{I}_{l}^{*}\left(\cmr\right)
\label{Eq:LensMap}
\end{equation}
 where summation is over each lens along the line of sight, $D_{l}$
is the distance factor, $\mathbf{I}_{l}\left(\cmr\right)$ is the complex 
scattering
function and the asterisk denotes complex conjugation (see 
\citet{Bourassa_Kantowski:1973}). In the case of spherically symmetric 
potentials, equation (\ref{Eq:LensMap})
simplifies to: 
\begin{equation}
\tilde{\cmr}=\cmr-\sum _{l}D_{l}\: 
\frac{\tilde{m}\left(\left|\cmr-\cmr_{l}\right|\right)\left(\cmr-\cmr_{l}\right)}
{\left|\cmr-\cmr_{l}\right|^{2}}
\label{Eq:LensMapSph}
\end{equation}
where $\tilde{m}\left(r\right)$ is the cylindrical mass, defined
as the total mass contained in a cylinder centered on the lens and
with radius $r$ equal to the impact parameter \citep{Dyer:1977:OsGl}. 
Depending
on the structure of the mass profiles and the magnitude of $D_{l}$,
equations (\ref{Eq:LensMap}) and (\ref{Eq:LensMapSph}) will have
either one or an odd number of solutions, corresponding to the strong
and weak lensing regimes.

Using equations (\ref{Eq:etag}) and (\ref{Eq:LensMap}), one can formalize
the value of \etag for an ideal jet (i.e., 
$\kappa \left(\theta \right)=0$).
Let $\Ps \left(\theta \right)$ be the tangent vector to the path
taken by the jet in the image plane, defined by $\Gam \left(\theta \right)$
($\Ps \left(\theta \right)=d\Gam /d\theta $).
Likewise, let $\tilde{\Ps }\left(\theta \right)$ and 
$\tilde{\Gam }\left(\theta \right)$
be the tangent vector and the path in the source plane (see \fig
\ref{fig:etag_defined}). Following equation
(\ref{Eq:LensMap}), we can write
\begin{equation}
\tilde{\Gam }\left(\theta \right)=\Gam \left(\theta \right)-
\sum _{l}D_{l}\: \mathbf{I}_{l}^{*}\left(\Gam \left(\theta \right)\right)
\label{Eq:tangent_lensed}
\end{equation}
We assume
that $\chi \left(\theta \right)=\tilde{\psi}\left(\theta \right) \equiv 
\arg(\tilde{\Ps})$
(i.e., perfect alignment between morphology and intrinsic polarization) and
by differentiating equation (\ref{Eq:tangent_lensed}), we obtain
\begin{equation}
\tilde{\Ps }\left(\theta \right)=\Ps \left(\theta \right)-\sum _{l}D_{l}\, 
   \left[\frac{\partial \mathbf{I}_{l}^{*}}{\partial x}\frac{dx}{d\theta }+
   \frac{\partial \mathbf{I}_{l}^{*}}{\partial y}\frac{dy}{d\theta }\right]
\end{equation}
where we have taken 
$\Gam \left(\theta \right)=x\left(\theta \right)+i\, y\left(\theta \right)$.
With this notation, the tangent vector can be written as 
$\Ps\left(\theta \right)=dx/d\theta + i\,dy/d\theta$ and the tangent angle
is $\psi = \arg\left(\Ps\right)$.
Substituting this into equation (\ref{Eq:etag}), we get the result:
\begin{eqnarray}
  \eta_{G}\left(\theta\right) & = &
  \arg\left\{\frac{dx}{d\theta} - \sum_l D_l \left(
  \frac{\partial Re\left(\mathbf{I}\right)}{\partial x}\frac{dx}{d\theta} +
  \frac{\partial Re\left(\mathbf{I}\right)}{\partial y}\frac{dy}{d\theta}\right)
  \right.\nonumber \\
   & & \left.  +i\frac{dy}{d\theta} + i \sum_l D_l \left(
  \frac{\partial Im\left(\mathbf{I}\right)}{\partial x}\frac{dx}{d\theta} +
  \frac{\partial Im\left(\mathbf{I}\right)}{\partial y}\frac{dy}{d\theta}\right)
  \right\} - \psi
  \label{eq:etag_comp}
\end{eqnarray}
Switching to component notation, equation (\ref{eq:etag_comp}) can be cast
in a more compact form:
\begin{equation}
   \eta _{G}\left(\theta \right)= \arg\left\{\sigma ^{ij}\left(\theta \right)\, 
   \psi _{j}\left(\theta \right)\right\} - \psi
   \label{Eq:etag_theo}
\end{equation}
were $\psi _{j}\left(\theta \right)=\left(dx/d\theta ,\: dy/d\theta \right)$
and $\sigma^{ij}$ is the shear matrix:
\begin{equation}
   \sigma ^{ij}=\left[
      \begin{array}{cc}
      1-\sum_l D_l\frac{\partial Re\left(\mathbf{I}\right)}{\partial x} & 
         -\sum_l D_l \frac{\partial Re\left(\mathbf{I}\right)}{\partial y}\\
      \sum_l D_l \frac{\partial Im\left(\mathbf{I}\right)}{\partial x} & 
         1 + \sum_l D_l \frac{\partial Im\left(\mathbf{I}\right)}{\partial y}
      \end{array} \right]
   \label{Eq:sigma}
\end{equation}
From this, we see that \etag is a measure of the differential
bending projected along the tangent vector of the jet. From equation 
(\ref{Eq:etag_theo}), we see that \(\etag = 0\) if \(\psi_j\) is an
eigenvector of \(\sigma^{ij}\).  In the case of a single, spherically
symmetric lens, this corresponds to radial and tangential vectors.
We can therefore imagine
pathological cases where a gravitational lens may be present, yet
one does not detect any signal. For example, in the case where the
jet follows a straight, radial path away from the lens, the differential
bending is parallel to the tangent vector everywhere on the jet and
\etag is therefore zero. Another, more unlikely case is where
the jet ``circles'' a spherically symmetric lens at a constant
impact parameter. In this case, the differential bending is zero along
the jet and \etag is again zero. 

\section{SIMULATIONS}
\label{sect:sim}
The purpose of simulating the weak lensing of an ideal jet is twofold:
(1) to determine the best way to objectively choose the fiducial
line and (2) to
test the sensitivity of \etag to the various
lens parameters. For our testbed, we
construct a fake jet 15\arcsec\ long by convolving a straight line with a 
Gaussian beam
having FWHM of 3\arcsec, which is typical of our observations.  This fake
jet can then be lensed with any mass distribution desired.  We choose
a spherically symmetric King profile, whose density is given by
\begin{eqnarray}
\rho (r) & = & \rho _{o}\left( 1+(r/a)^{2}\right) ^{-3/2},\; r\leq r_{c}\label{Eq:rho_King_ell} \\
\rho (r) & = & 0,\; r \ga r_c\nonumber 
\end{eqnarray}
where $a$ is the core-radius, the radius at which the density falls to about 
35\% of its central value ($\rho_o)$ and $r_c$ is the cut radius.  Unless 
otherwise
stated, all the modeling in this paper will be done with this potential.
We use this
lensed jet to first test our method for determining the fiducial line,
$\Gam\left(\theta\right)$.

\begin{figure}
\plotone{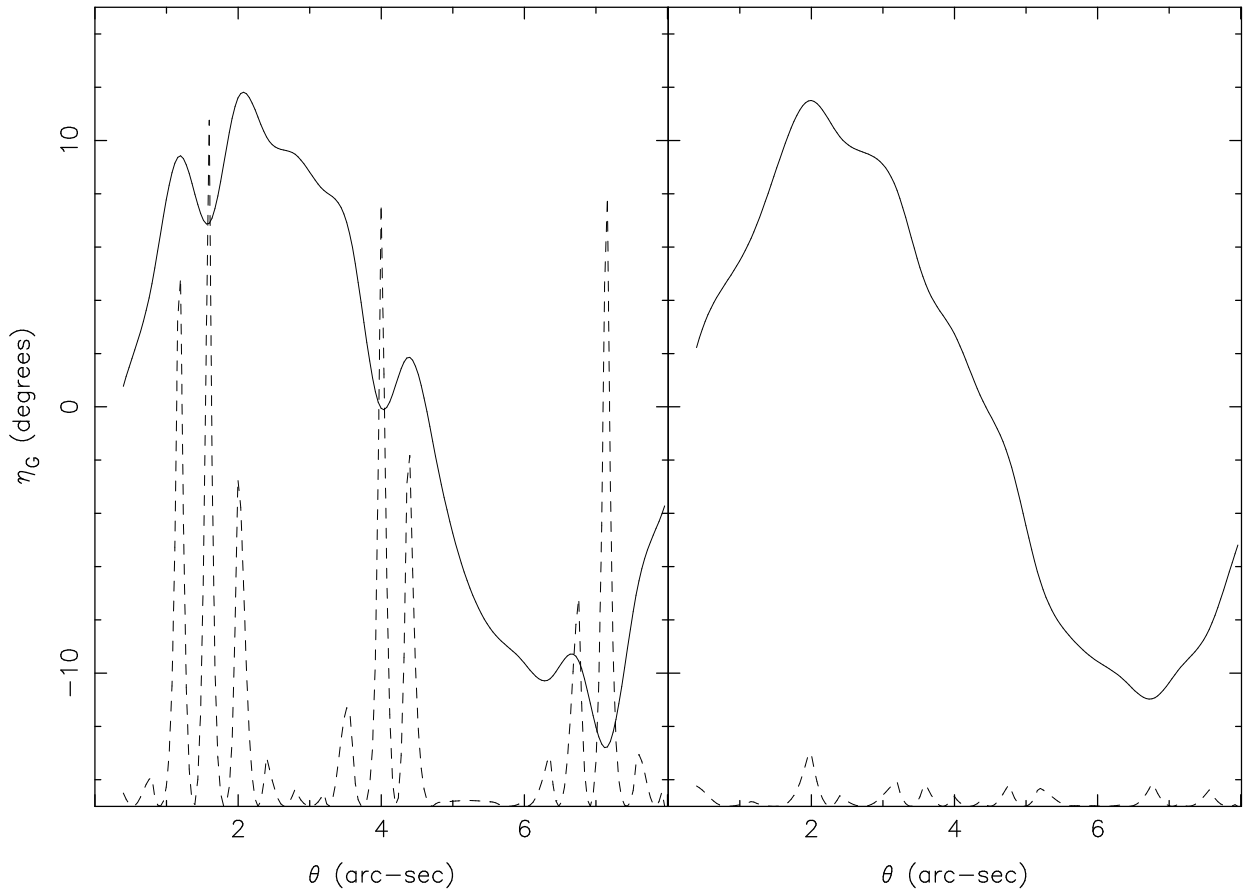}
\figcaption{Plot of \etag using only the contour averaging algorithm
(first panel)
and the contour averaging followed by the curvature minimization algorithm
after 7 iterations (second panel). In both cases, the dashed line is 
proportional
to the curvature (or inversely proportional to the radius
of curvature) of \etag along the jet. The horizontal axis represents
the distance along the jet in arc-seconds. \label{Fig:etag_curvature}}
\end{figure}

Initial values of $\Gam\left(\theta\right)$ 
are computed by taking a weighted average of the jet's 
isophotal contours. 
This typically results in an \etag signal which
shows the correct sinusoidal shape as well as
the expected small scale random variations due to badly chosen jet points
\citep{Kronberg:96}.  The \etag signal has a scale comparable to the
angular scale of the lens itself and therefore has
a large radius of curvature.  Small-scale variations 
represent effects that are not gravitationally induced
and are recognized by their small
radii
of curvature.  To correct these errors, the jet points
responsible for the variations are located and varied by a small amount 
perpendicular to the jet axis in an attempt to reduce the small-scale 
curvature of the 
\etag signal.  The gravitationally induced \etag signal will not be 
eliminated by
varying the position of just one jet point.
\fig \ref{Fig:etag_curvature} shows the measured value of \etag
for a lensed jet before and after this correction is performed. 

We now proceed
to model the weak lensing and extract the parameters.  The downhill simplex 
method \citep{Press:92} is used to find the best fit by minimizing
the $\chi ^{2}$ statistic. Assuming that the angular position
and redshift of each lens has been measured, there
are only 4 remaining parameters to be fitted: the total mass $M_{G}$
probed by the lens, the mass profile function $\rho\left(r\right)$, the core 
radius
$a$ and the cut radius $r_{c}$.
One should be mindful of the shape of the parameter space
and we now turn to simulations of $\chi ^{2}$ to attempt to ascertain
the degeneracies in the parameters. 

\begin{figure}
\plotone{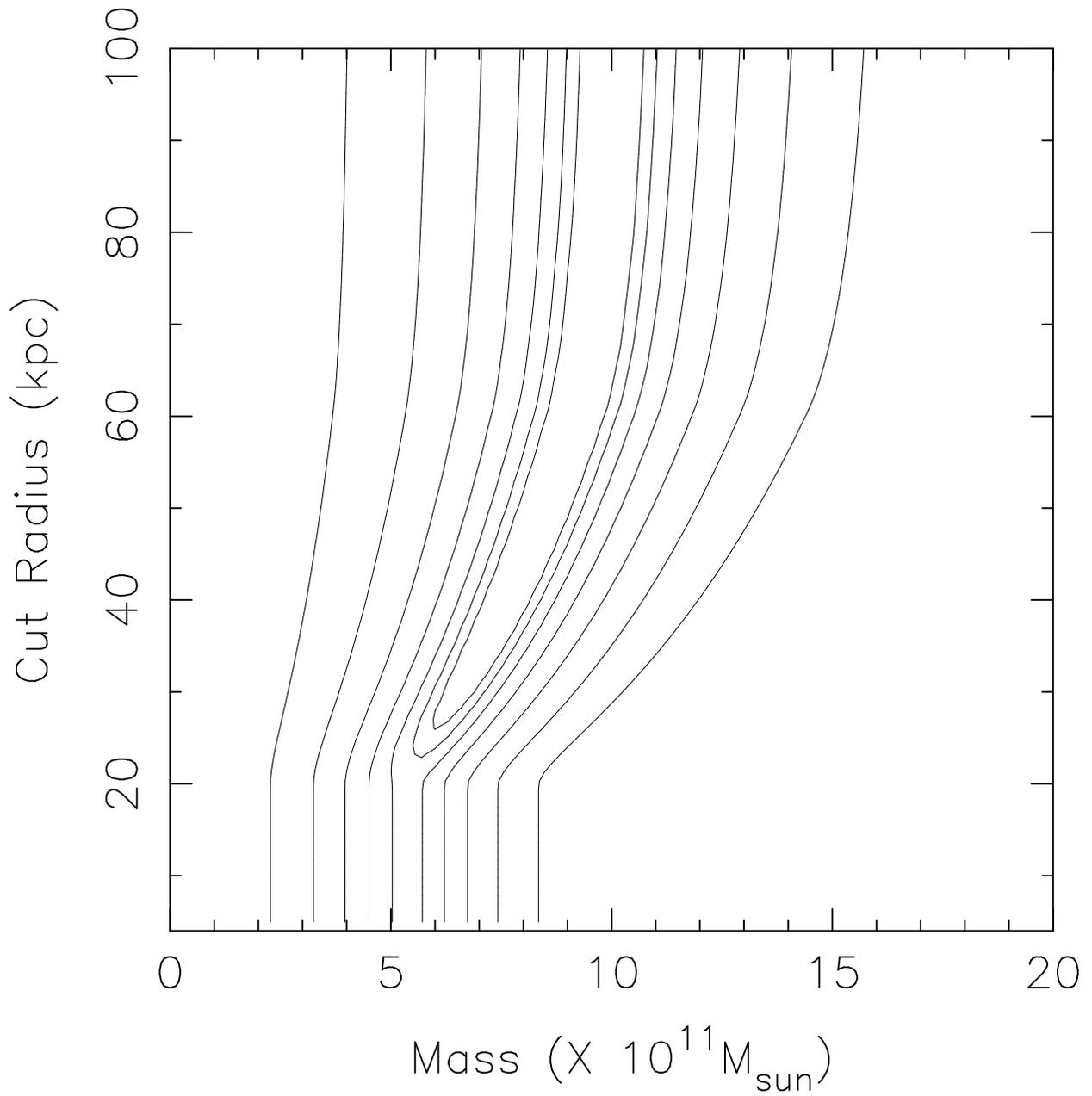}
\figcaption{$\chi ^{2}$ as a function of the mass $M_{o}$ (in $\mgal $) and
the cut-off radius $r_{c}$ in $\kpc $ for the King galaxy profile. 
The solid contours denote $\chi ^{2}=5,10,20,40,60,120$ and $240$.\label{Fig:chi_sq.mass_cut}}
\end{figure}

We begin with the cutoff radius,
$r_{c}$. This is only relevant to the mass models whose profiles do not
converge to a finite mass as $r \rightarrow \infty$.
One might expect that as the cutoff radius is increased,
the best fit value for the total mass will increase so
as to maintain enough mass at smaller radii to reproduce the observed
lensing. \fig \ref{Fig:chi_sq.mass_cut} shows the value of $\chi ^{2}$
as a function of the model mass and cut-off radius. Indeed, in the middle impact
parameter zone
($20 \la r_c \la 60\ \kpc $), larger
cut-off radii require larger masses. We note however that $\chi ^{2}$ becomes
independent of $r_{c}$ for $r_{c}\la20\ \kpc $ because the cut-off radius
is now less than the smallest impact parameter, which we shall
denote as $R_i$ of the lens-jet system (see \fig \ref{fig:etag_defined}).
It also becomes less sensitive to $r_c$ for $r_{c}\ga60\ \kpc $, which is 
comparable to the
furthest impact parameter ($R_o$) for the jet-lens system. This is due to
the fact that as the cut-off radius is increased, the additional mass 
outside $R_{o}$ acts more
like a mass-sheet, to which \etag (and any other weak-lensing analysis) is 
increasingly insensitive.
It is therefore not worth investigating values of $r_{c}\gg R_{o}$.

\begin{figure}
\plotone{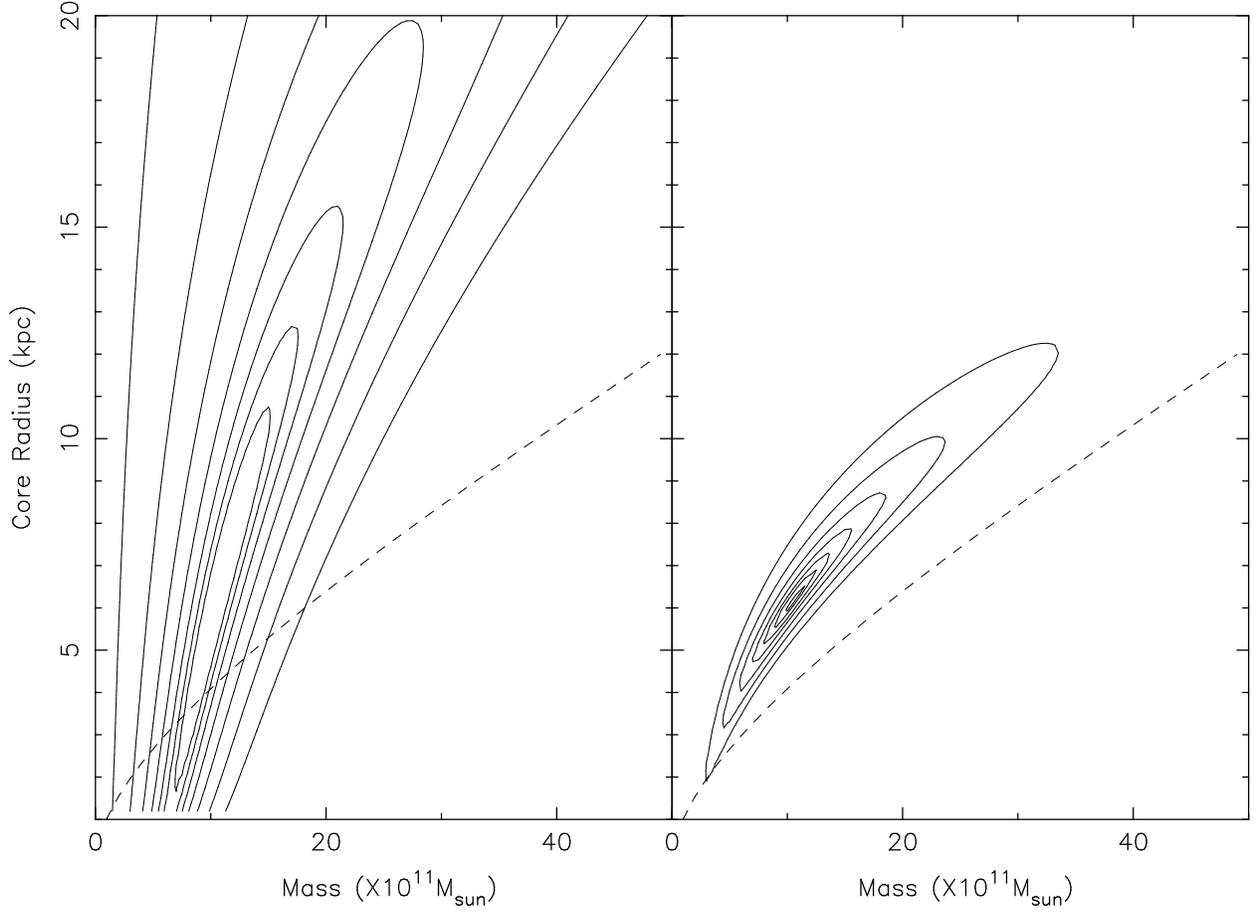}
\figcaption{$\chi ^{2}$ as a function of the mass $M_{o}$ (in $\mgal $) enclosed
within a radius $R_{o}=60\ \kpc $ and the core radius $a$ in $\kpc $
for the King profile. The solid contours denote $\chi ^{2}=5,10,20,40,60,120$
and $240$. The dashed line shows the minimum core radius such that
there is no multiple imaging. The lens is placed at $5\arcsec $ (first panel)
and $1\arcsec $ (second panel) from the center of the jet. \label{Fig:chi_square_far}}
\end{figure}

\fig \ref{Fig:chi_square_far} shows the value of $\chi ^{2}$ as
a function of the mass and core radius for a King profile lens at
two different distances from the jet.
Clearly, in the first case, the galaxy's mass is well constrained by the 
modeling; however, the
core radius is not so well constrained. This is because the jet probes
radii ranging from $R_{i}=20\ \kpc $ to $R_{o}=60\ \kpc $ and is therefore
relatively insensitive to changes in the internal structure (i.e.,
$R \la 20\ \kpc $) of the lens. If one also imposes the constraint that
the lens be incapable of producing multiple images, then
a lower limit can be placed on the core radius, as shown by the dashed
line in the first panel of \fig \ref{Fig:chi_square_far}.
If the jet probes smaller radii
of the lens potential, the modeling will be able to better constrain
the core radius, as shown the second panel of \fig \ref{Fig:chi_square_far},
where
the lens has been moved closer (1 \arcsec) to the jet such that $R_{i}=10\ \kpc $.

\begin{figure}
\plotone{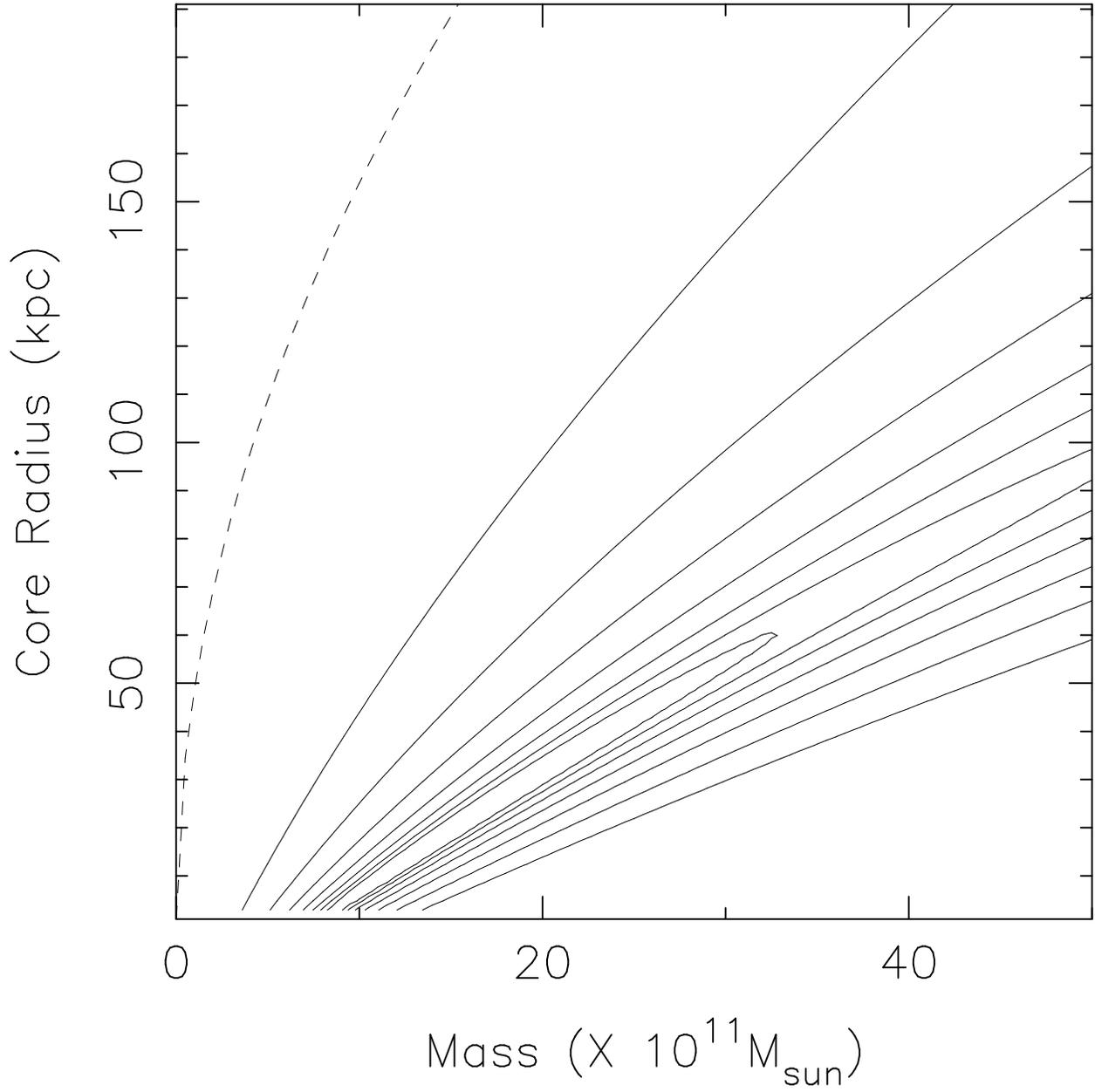}
\figcaption{Same as \fig \ref{Fig:chi_square_far}, except that a deVaucouleurs
mass profile is used instead of a King profile. Note that all the
$\chi ^{2}$ contours are excluded by the single-image condition (dashed
line).\label{Fig:chi_sq.gdev}}
\end{figure}

The mass profile function is usually not well constrained by the modeling, except in 
extreme cases
such as the de Vaucouleurs exponential profile where the
core is sharp enough that one cannot get enough mass interior to $R_{i}$
and $R_{o}$ without causing the jet to be multiply imaged. This case is
shown in \fig \ref{Fig:chi_sq.gdev}.
Note that all the $\chi ^{2}$ contours lie under the minimum core
radius curve (dashed line). 

\section{MODELING OF 3C9}
\begin{figure}
\plottwo{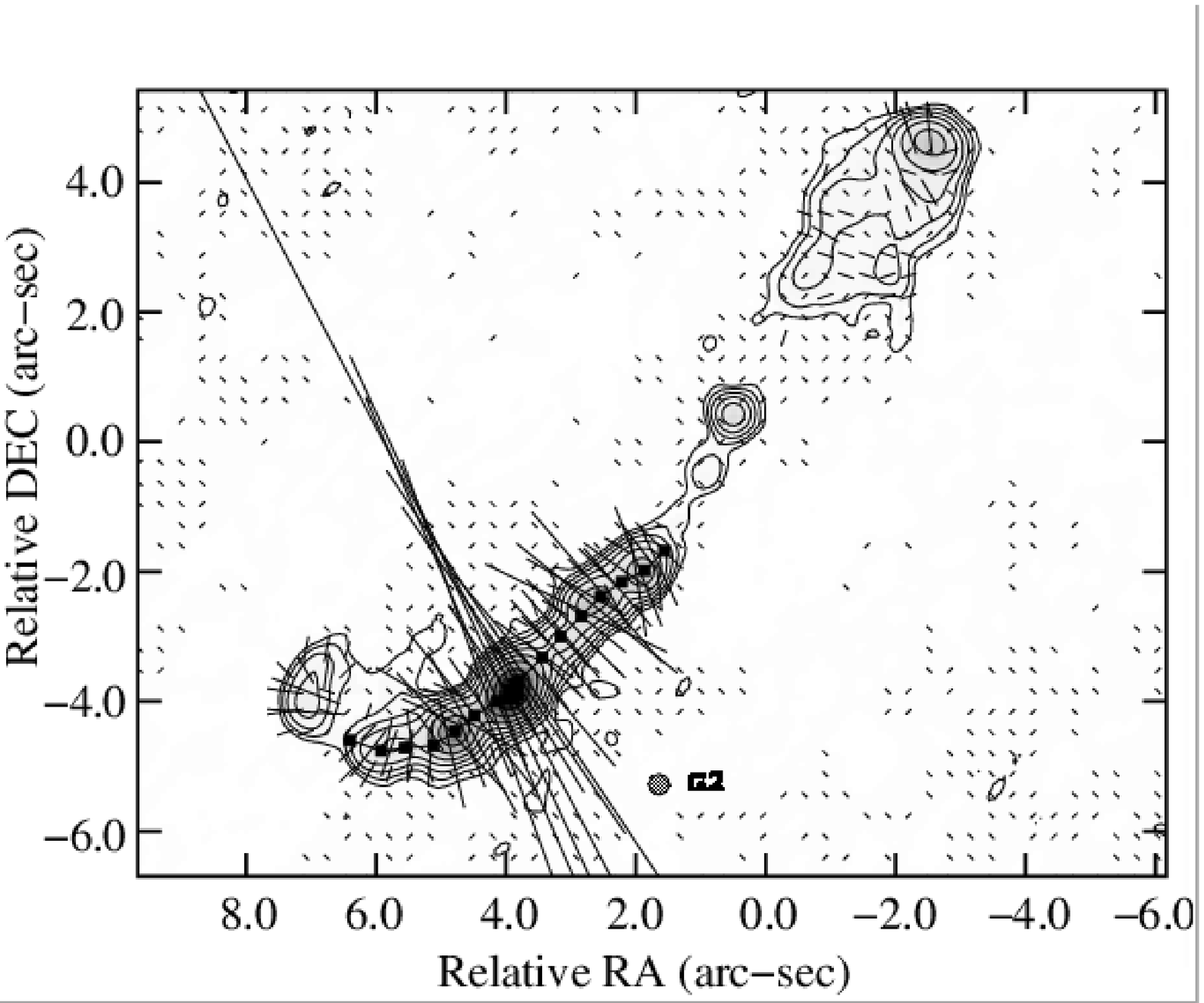}{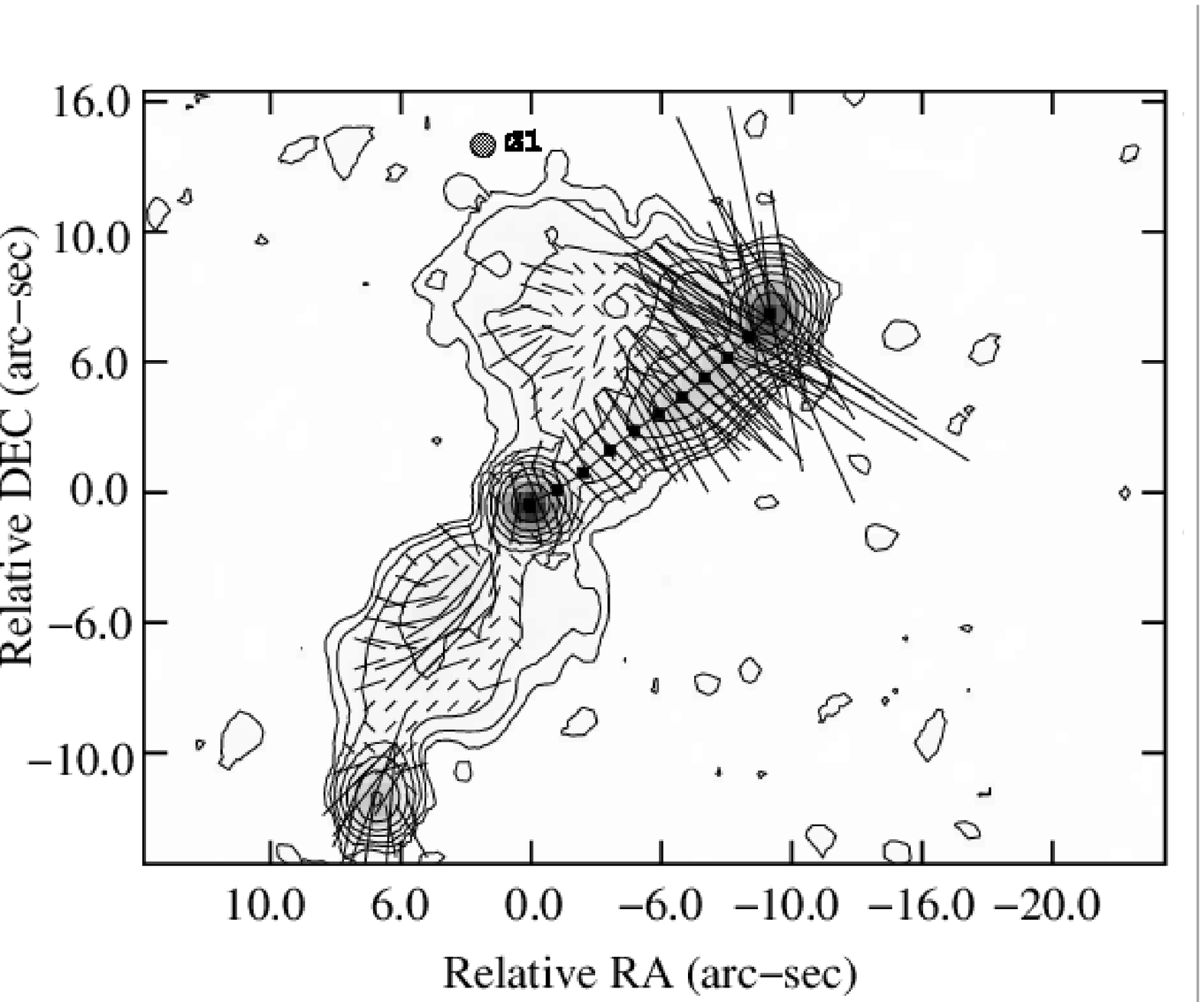}
\figcaption{Radio contour map of quasar 3C9 at X-band and 1253+104 at C-band.
Contours are drawn at 0.12, 0.25,
0.5, 1, 3, 6, 12, 25 and 50 percent 
of peak intensity, which is \( 56.5\ \mathrm{mJy}\, \mathrm{beam}^{-1} \) for
3c9 and \( 29.7\ \mathrm{mJy}\, \mathrm{beam}^{-1} \).
The magnetic field components of the polarization are drawn as vectors
such that \( 1\arcsec  \) is equivalent to a polarization
intensity of \( 1\mathrm{mJy}\, \mathrm{beam}^{-1} \).
The filled boxes connected by lines represent the fiducial line. 
\label{fig:3c9+1253}}
\end{figure}

The quasar 3C9 (\( z=2.012 \)) was the first object analyzed with
the alignment breaking technique. The first panel of
\fig \ref{fig:3c9+1253} shows the radio emission and polarization vectors
for 3C9 as well as the
the fiducial line defining the jet's morphology.  For clarity, we plot
the magnetic field component of the polarization, which is simply rotated
\(90 \degr\) with respect to the electric field component\footnote{%
In their original definition of \(\etag\), \citet{Kronberg:91}
explicitly subtract \(90\degr\) from the observed polarization for this
reason.}.  It is an 
interesting case in that
initially, only one lens (a galaxy at \( z=0.2538 \)) was in evidence
\citep[see][]{Kronberg:91}. However, the impact parameter and tangential 
position
made it difficult to model the observed \( \eta _{G} \) with 
only one galaxy and this led the authors to obtain deeper images of the field.
The result was the detection of a fainter galaxy with a smaller
impact parameter \citep[see][]{Kronberg:96}. While most 
likely less massive, the latter galaxy provides a better fit to the observed
values of \( \eta _{G} \) due to its small impact parameter. The
fact that the fainter galaxy was first detected gravitationally demonstrates
the power of our technique.  Unfortunately,
it was not possible to directly obtain a 
spectroscopic redshift for the fainter galaxy.

\begin{figure}
\plotone{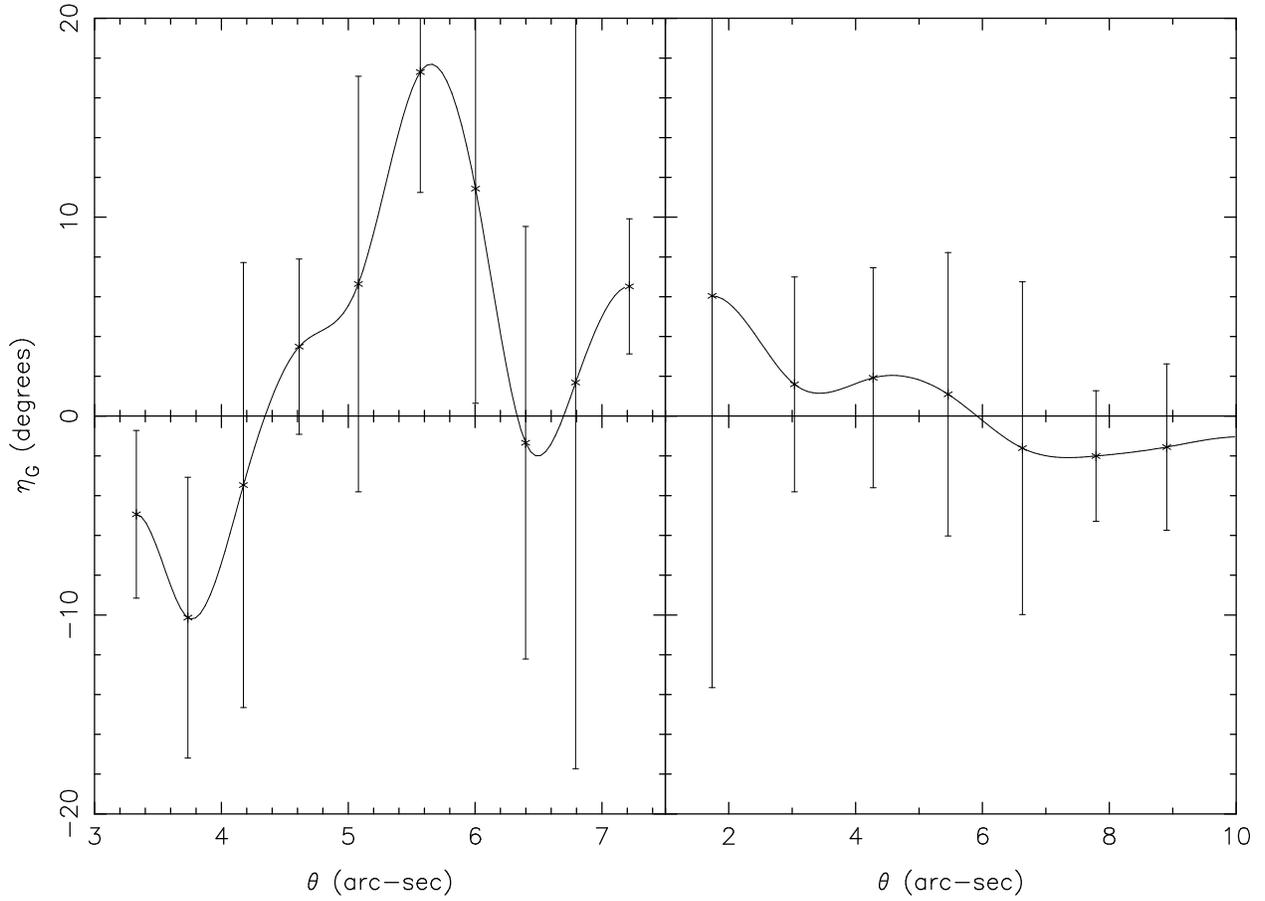}
\figcaption{\( \eta _{G}\) for the quasar 3C9 (first panel) and QSO 1253+104
(second panel). The horizontal
axis shows \(\theta\), the distance along the
jet, in arc-seconds from the quasar center. The vertical axis shows
\( \eta _{G}\) in degrees.\label{Fig:etag.3c9}}
\end{figure}

The plot for the observed values of \( \eta_{G} \) along the jet is shown 
in \fig
\ref{Fig:etag.3c9}. The error bars are computed by adding the intrinsic
scatter of the polarization
and the uncertainty in choosing the
jet points, which are determined by varying their positions 
by \(0.1\arcsec  \) perpendicular to the axis of the jet.  There is a clear 
variation of \( \eta _{G} \), as is expected
from the location of the lenses.

\begin{figure}
\plottwo{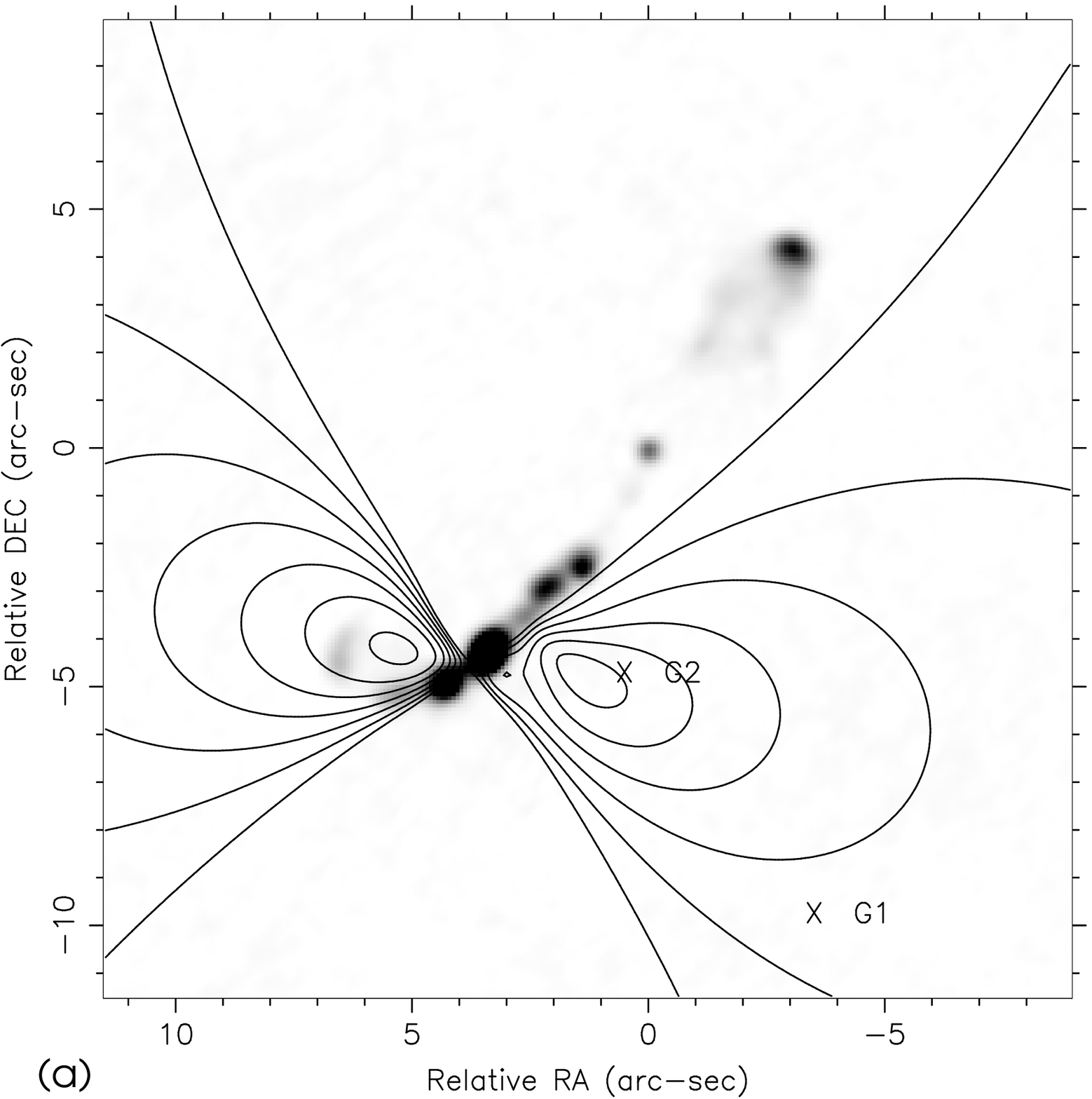}{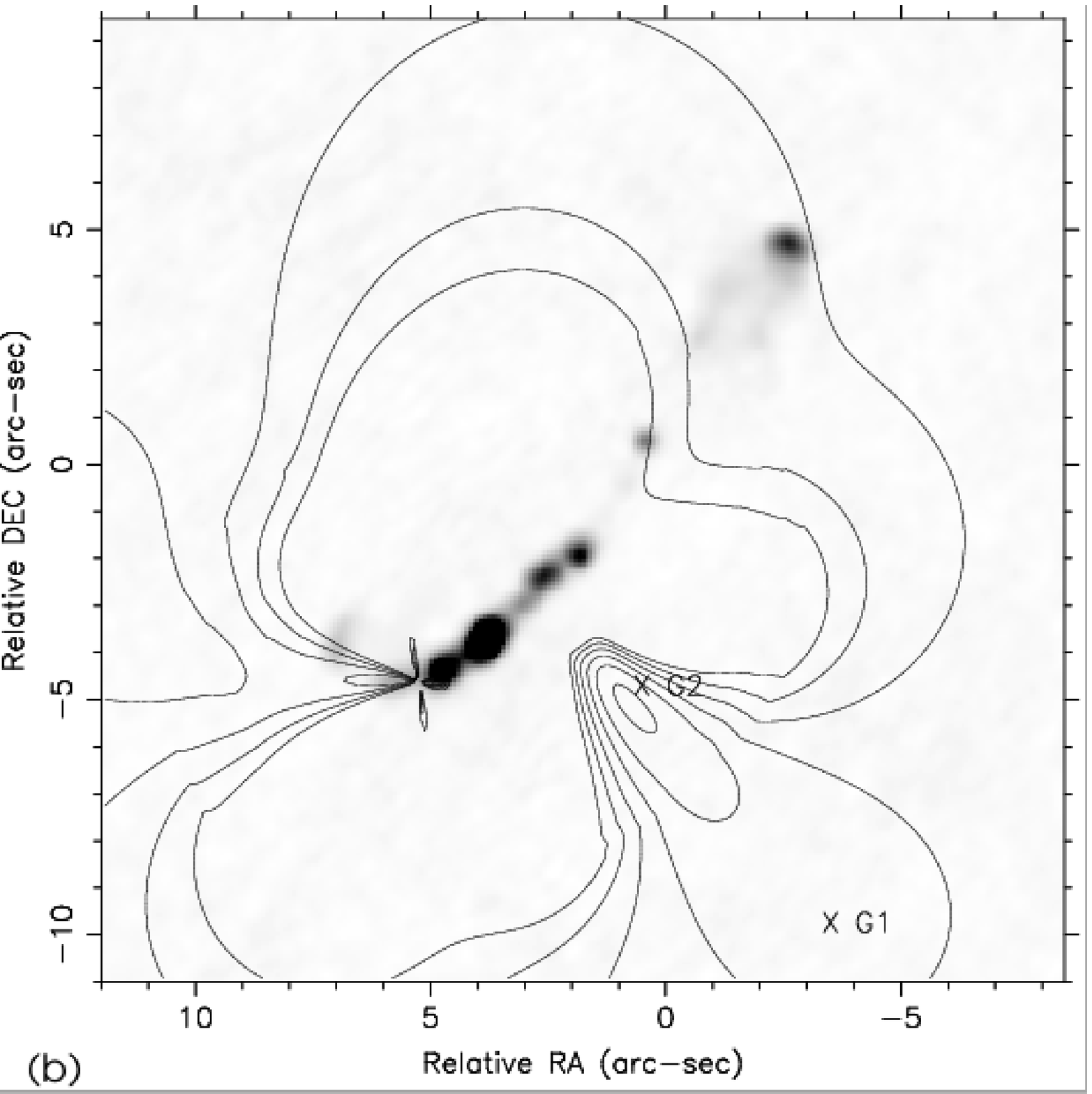}
\figcaption{\label{Fig:3c9_chi_x_y}\( \chi ^{2}\) as a function
of lens position on the sky for two different lens masses. The gray-scale 
image is of the quasar and
jet, while the contours are of \( \chi ^{2}\). (a) The lens mass is 
\(10 \times 10^{11} M_\odot\) and 
contours are drawn at \( \chi ^{2}= \)50, 100, 150,
200, 250, 300 and 350. (b) The lens mass is \(30 \times 10^{11} M_\odot\) 
and contours are drawn at \(\chi^2 = \)5, 10, 15, 20, 25 and 30.  Galaxies G1 
and G2 (as denoted in \citep{Kronberg:96})
are labeled for reference.  In both cases, the lens redshift is 1.0,
the core radius is 2 \kpc\ and the cut radius is 30 \kpc.}
\end{figure}

Even though the lens positions have been measured, it is instructive
to measure \( \chi ^{2} \) as a function of lens position
to see whether or
not the \( \eta _{G} \) signal itself points to the likely location
of the lens in the absence of a detection of the galaxies. We choose a King 
profile with mass \( 10\times 10^{11}M_{\odot }\)
and ensure that there are no multiple images produced, consistent with
observations. We then compute
\( \chi ^{2} \) for putative positions of a single lens over a search area 
on the sky.
Figure \ref{Fig:3c9_chi_x_y}a
shows the results. Note that there are two preferred locations for
minimizing \( \chi ^{2} \): to the east and to the west. In this case,
the best fit actually occurs to the east of the quasar jet, whereas
the observed lens is to the west, close to the second minimum.  Both these
positions are favored because they tend to straighten the jet in the source
plane.  However, if we increase the mass, the minimum \(\chi^2\) shifts to
the west and the eastern minimum all but disappears.  This is shown in \fig
\ref{Fig:3c9_chi_x_y}b.  The reason the minimum \(\chi^2\) does not correspond
exactly to
the location of G2 is because it is sensitive to sum of the masses of both
lenses and G1 in effect ``drags'' the minimum a little to the east.

\begin{figure}
\plottwo{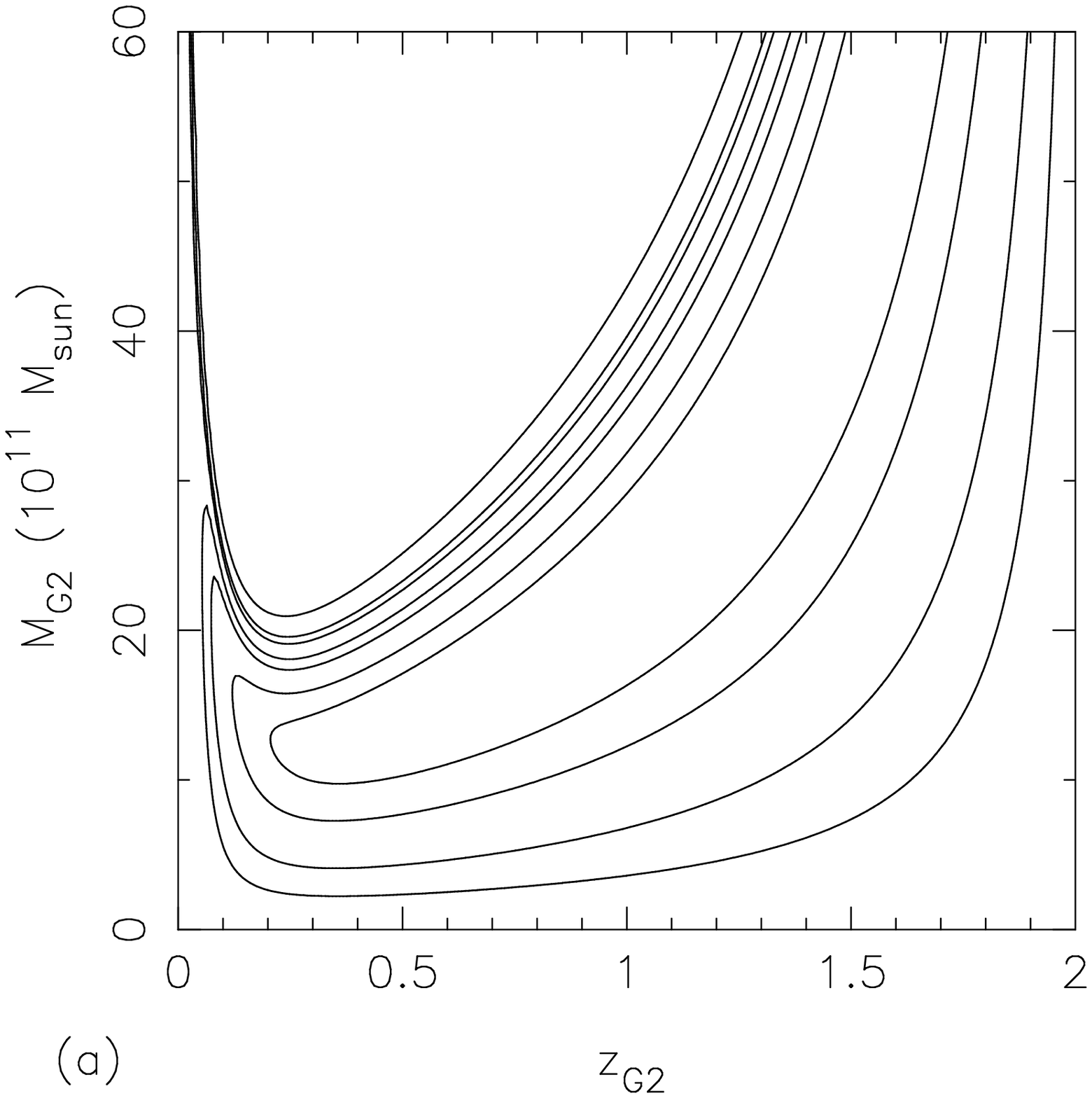}{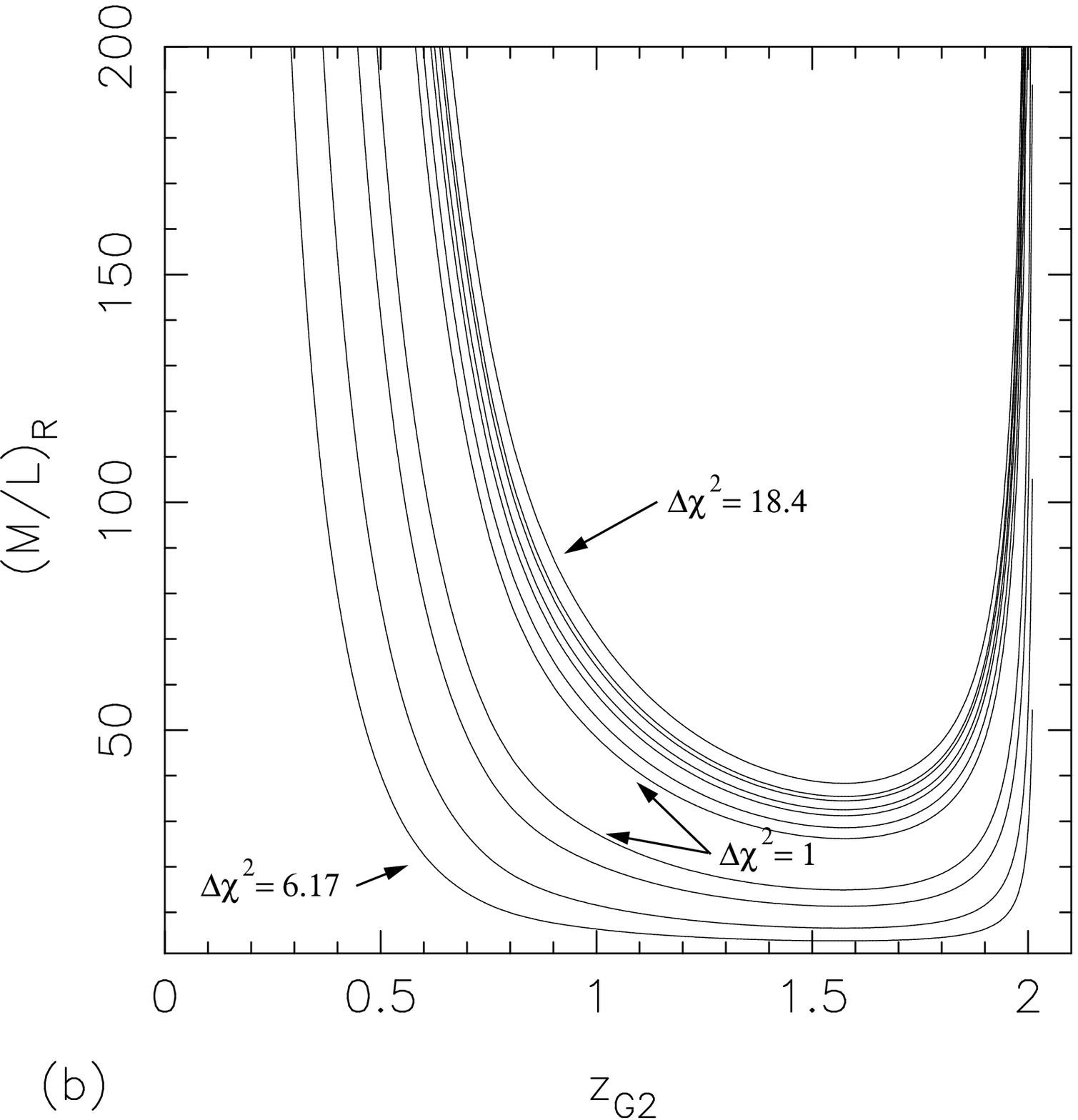}
\figcaption{\label{Fig:3c9_chi_z_m_ML}(a) \(\chi^2\) as a function of redshift
and mass of lens G2.  (b) \(\chi^2\) as a function of redshift and \ML\ for
lens G2.  The minimum $\chi^2$ is $3.6$ and contours are drawn at 
\(\Delta \chi^2=1.0, 2.3, 4.6, 6.17, 9.21, 11.0\) and $18.4$.  In both cases
, the core and cut radii are 5 \kpc\ and 30 \kpc, respectively, and
the mass of G1 is held at \(17\times 10^{11} M_\odot\).}
\end{figure}

Given that the positions of the two lenses in the field of 3C9
are observationally determined, we can
allow the other parameters to vary in this ``galaxy mass laboratory''. An
obvious combination to explore is the masses of both lenses. We therefore 
compute
\( \chi ^{2} \) for different combinations of masses for galaxy 1
and 2 (G1 and G2). Because G2 has no measured redshift, 
we cannot provide an upper bound to its mass.  However, 
a lower limit to its mass is determined by fixing its redshift at 
\( z_{2} \simeq 0.25 \). This corresponds to the maximum of the distance
factor.  However, since the computed absolute luminosity of the galaxy is also
a function of redshift, \(z_2 = 0.25\) does not yield the minimum
\ML\ (it would have to be in excess of several thousand \(M_\odot/L_\odot\)). 
Given this uncertainty, we compute \(\chi^2\) letting
the redshift and mass vary to find the minimum mass of G2 consistent with the
data. 
The results are shown in \fig \ref{Fig:3c9_chi_z_m_ML}a. 
We then do the same exercise, but now letting the redshift and \ML\ for galaxy 2
vary, the results
of which are shown in \fig \ref{Fig:3c9_chi_z_m_ML}b.  Here, we have used an
apparent magnitude of
\(m_R = 23\) \citep{Kronberg:96}. K-corrections for an Sbc galaxy are given in 
\citet{Coleman:1980},
and we assume an Einstein-deSitter cosmology\footnote{%
We choose this cosmological model in order to be consistent with 
\citet{Kronberg:96} with \(H_o = 75\).  For the redshifts of interest in this paper, the
results are quite insensitive to $\Omega_m$ and $\Omega_\Lambda$.}.
As predicted, the minimum mass occurs at \(z \simeq 0.25\) and is
approximately \(10 \times 10^{11}M_{\odot} \).  However, examining the
behavior of \ML\ with redshift, a low \ML\ is favored by a lens with a redshift
in the range (\(1 \la z_2 \la 1.8\)), where its value  \(\ML \simeq 20\) is
largely insensitive to changes in redshift.  Interestingly, there is an
absorption feature in 3C9's spectrum at a redshift of \(z_{abs} = 1.6\)
\citep{Kronberg:96}.
This would, however, require a mass for G2 on the order of 
\(6\times 10^{12}M_{\odot}\) and R-band luminosity of 
\( 2\times10^{11}L_{\odot,R}\).  

The smallest impact parameter between the
jet and G2 is approximately 2 arc-seconds which at a redshift \(z\simeq 1\)
corresponds to a physical distance of 12 \kpc.  We therefore find that the
best fit \(\chi^2\) is insensitive to the core radius, as discussed in
\S \ref{sect:sim}.  We can, however, use the fact that there are no observed
secondary images of the quasar or jet to place lower limits on the core
radius of G2, provided its redshift is assumed.  For both \(z_2=1.0\) and 
\(z_2 = 1.6\), we establish a minimum core radius of 5 \kpc.

\begin{figure}
\plottwo{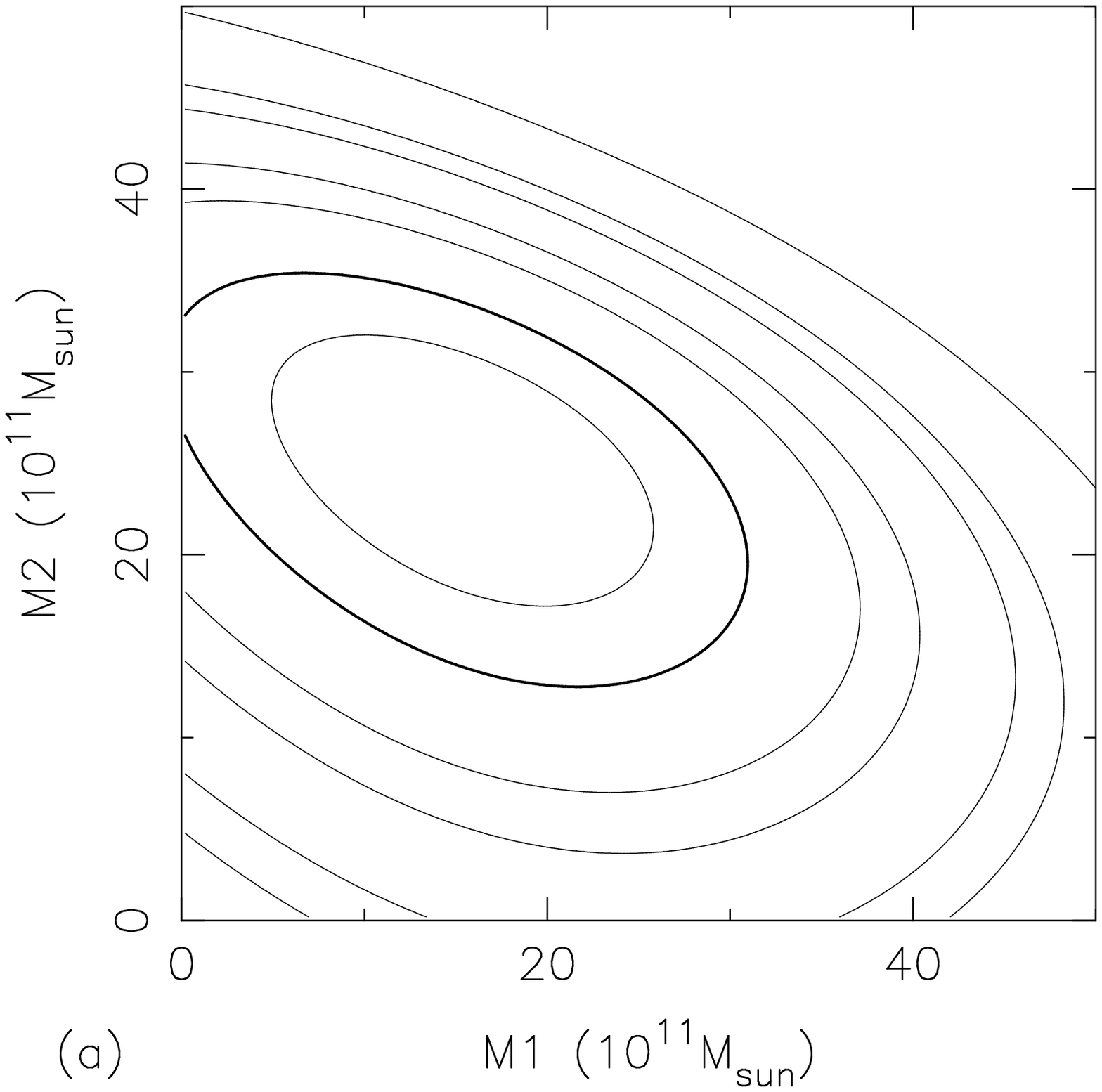}{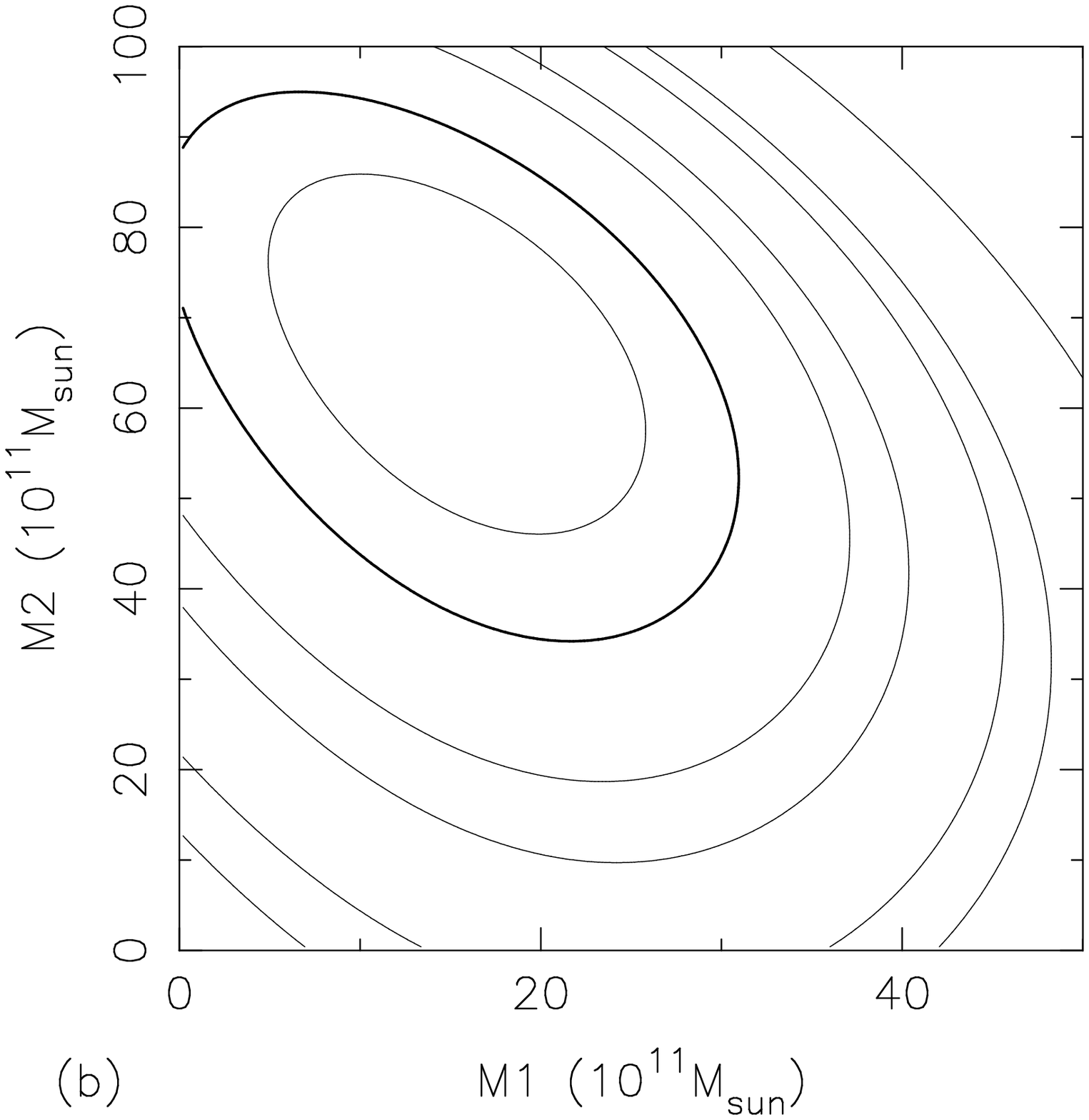}
\figcaption{\label{fig:3c9_chi_m1_m2_z}\(\chi ^2\) as a function of the 
masses of G1 and G2 for two
different redshifts of G2:  (a) \(z_2 = 1.0\) and (b) \(z_2 = 1.6\).  Contours
are drawn at the same levels as figure \ref{Fig:3c9_chi_z_m_ML}.  
The \(\Delta \chi^2 = 2.3\) contour is taken as the 1-\(\sigma\)
confidence region for the joint probability of G1 and G2 and is drawn as a 
bold line.  In both cases, core and cut radii for the two lenses are 5 \kpc\ 
and 30 \kpc, respectively.}
\end{figure}

Unlike G2, the redshift of G1 has been measured and one
can place confidence limits on its mass.  By varying the mass of G1 and G2
around the best fit parameters, we get the \(\chi^2\) contours shown in
\fig \ref{fig:3c9_chi_m1_m2_z}, where we have set \(z_2\) to be the two most likely redshifts
for G2 (\(z_2=1.0\) and \(z_2 = 1.6\)).
Clearly, the
mass of G2 is more tightly constrained than G1 (when the redshift is
fixed), since G2 is closer to the jet.  Note that the redshift and
mass of G2 do not affect the best estimate of G1 and so we find that
G1 is
``decoupled'' from the parameters of G2, as noted in \citet{Kronberg:96}.

\begin{figure}
\plotone{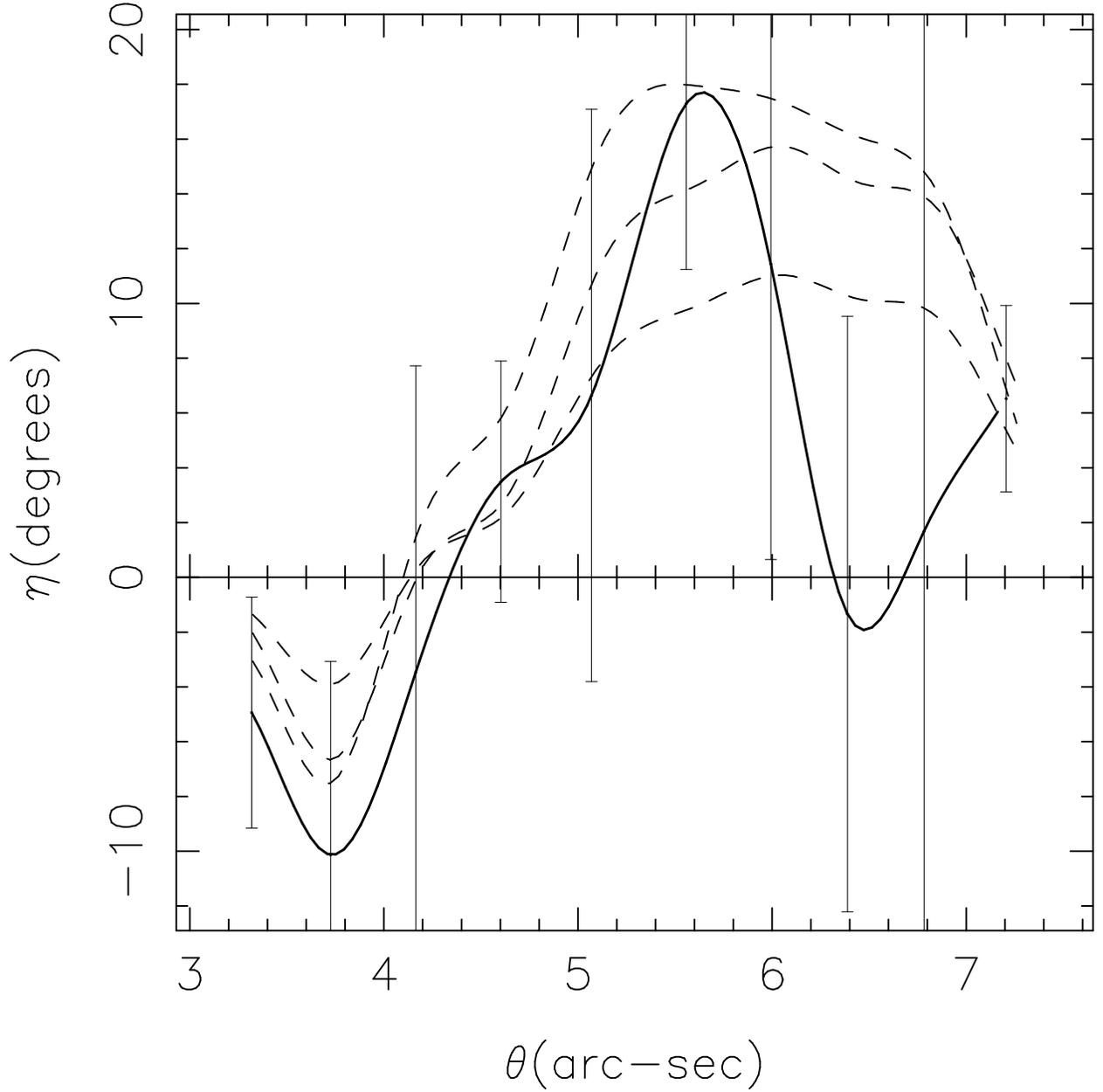}
\figcaption{\label{fig:etag_fits} Best fit \etag assuming \(z_2 = 1.0\),
\(M_{G2} = 3 \times 10^{12} M_\odot\) and 
\(M_{G1} = 1.7\times 10^{12} M_\odot\).  The observed data (solid line)
and three representative fits are plotted (dashed lines).  The middle dashed
curve is
the best fit and the two outer curves 
correspond to 1-\(\sigma\) fits.  
The core and cut radii are 5 \kpc\ and 30 \kpc, respectively, for both lenses.}

\end{figure}

Figure \ref{fig:etag_fits} shows the best fit as well as two typical
fits corresponding
to the \(1\sigma\) confidence interval.  None of our models can reproduce
the very sharp decrease of \etag at \(\theta > 6\arcsec\).  In this
particular system, the ``sharpness'' of the model \etag curve is
largely a function of the projected distance of the galaxy G2 from
the jet and is fixed by observation.  However, the region \(\theta > 6 \arcsec\)
is where the jet makes a sharp turn and rapidly decreases in
brightness (see Figure \ref{fig:3c9+1253}).  It may be that this is
the transition region between the jet and lobe, where the assumption of
alignment between jet morphology and polarization breaks down.  Nevertheless,
our models do reproduce the \etag behavior in the region 3 \(< \theta <\)
6 \arcsec, where the signal is highest and the jet morphology is
unambiguous.

Table \ref{tab:params} summarizes the parameters derived from the preceding
analysis.  For the purposes of comparison, we list the best fits for a
variety of redshifts for G2.  The last row gives results when no constraints
are placed on \(z_2\) and so should be taken as the firm limits.  In this case,
the only 
parameter which can be fully
constrained is the mass of G1 out to approximately 10 arc-seconds (\(r\simeq30\)
\kpc\ at \(z=0.25\)).  The confidence 
intervals were determined by projecting the \(1 \sigma\) \(\chi^2\) surface
onto the axis of the parameter in question.  For the joint probability
distribution of $M_{G1}$ and $M_{G2}$, this corresponds to $\Delta \chi^2=2.3$,
whereas the confidence interval for $M_{G1}$ only (last row of Table
\ref{tab:params}) corresponds to $\Delta \chi^2 = 1$.  The masses quoted for 
G2 are
out to approximately 6 arc-seconds, which corresponds to \(r\simeq 35\) \kpc\ at
redshifts \(z\simeq 1\).
Given that our estimates for
the mass of G2 at each redshift are for \(r \la 6\) arc-seconds, well beyond 
the observed light of G2, we give \(\left(\ML\right)_{G2}\) as upper limits.

\section{QSO1253+104: OBSERVATIONS}
\label{sect:Observations}
QSO1253+104 is a classic FRII radio source at a redshift of
\( z=0.83 \). The jets extend to the North-West and to the South-East
approximately \( 10\arcsec  \) in each direction. Both jets show
extended structure, especially the North-West jet/lobe. \fig 
\ref{fig:3c9+1253}
shows the radio contour map at 5 GHz. The data was taken
at the VLA as part of the \( \eta _{G} \) survey conducted by Kronberg
et al.
The first deep optical image of the field in the literature is due
to \citet{Hutchings:1992}, in which 3 objects are visible within 
\( 30\arcsec  \) of the QSO. 

\begin{figure}
\plotone{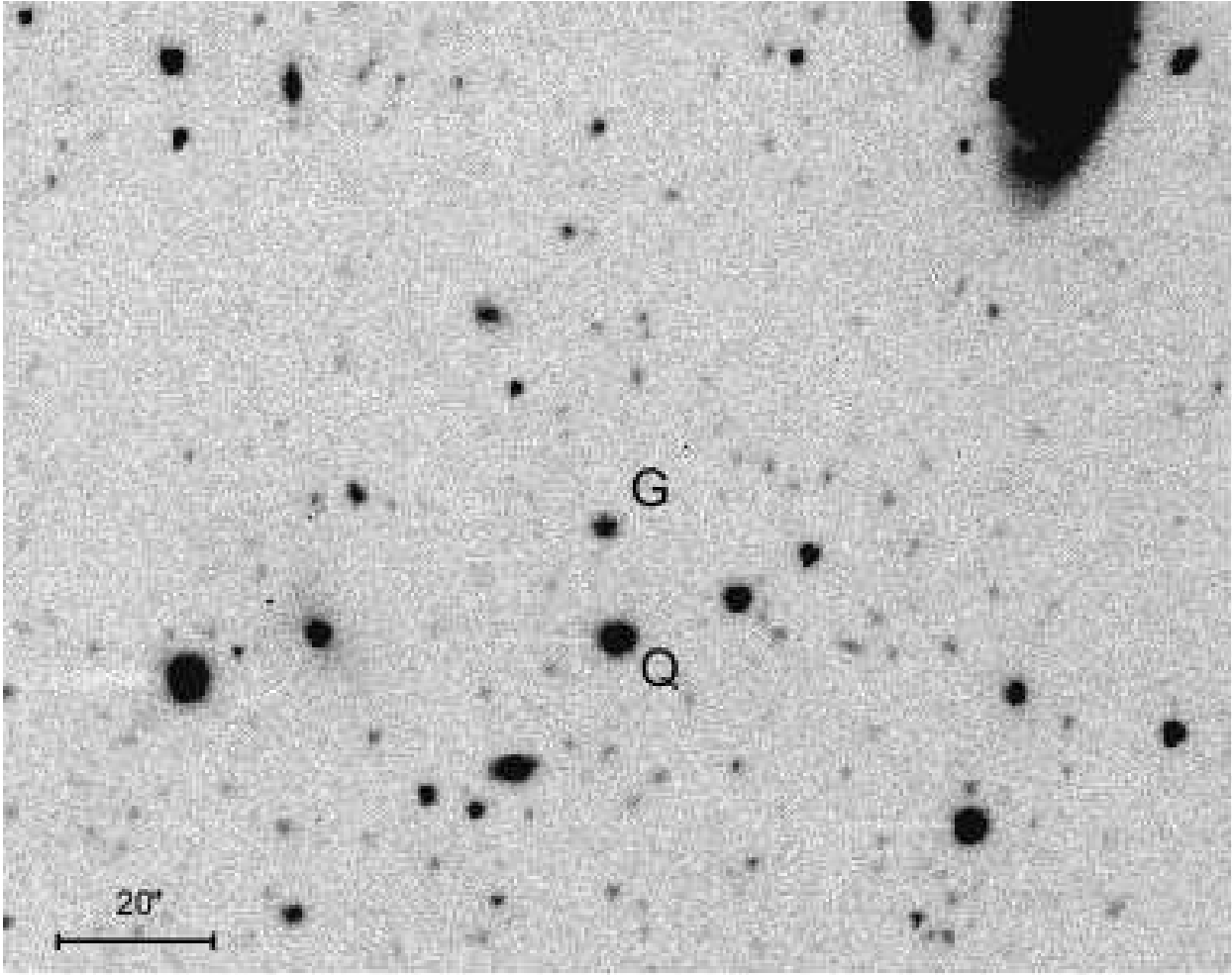}
\figcaption{
Sum of three 600\,sec exposures of the field of Q1253+104 in the R-band. The
quasar is labelled Q and the galaxy with G. North is up and East to the left.
\label{fig:1253_opt}}
\end{figure}

A sample of distant quasars with large scale jets was observed since June 1993
during several runs on the 3.5m-telescope on Calar Alto using FREDUK and MOSCA.
The vicinity of the jets was searched for intervening galaxies which
would cause ``alignment breaking''.
QSO1253+104 was imaged in B, R and I on 1997 February 10 with MOSCA using a
$2\mathrm{k}\times2\mathrm{k}$ Loral CCD with $15\,\mu$ pixels corresponding 
to 0.32\,arcsec on
the sky, following the earlier detection of a candidate galaxy with FREDUK in
June 1994. \fig \ref{fig:1253_opt} shows the sum of three 600\,sec images in 
the R-band.

\begin{figure}
\plotone{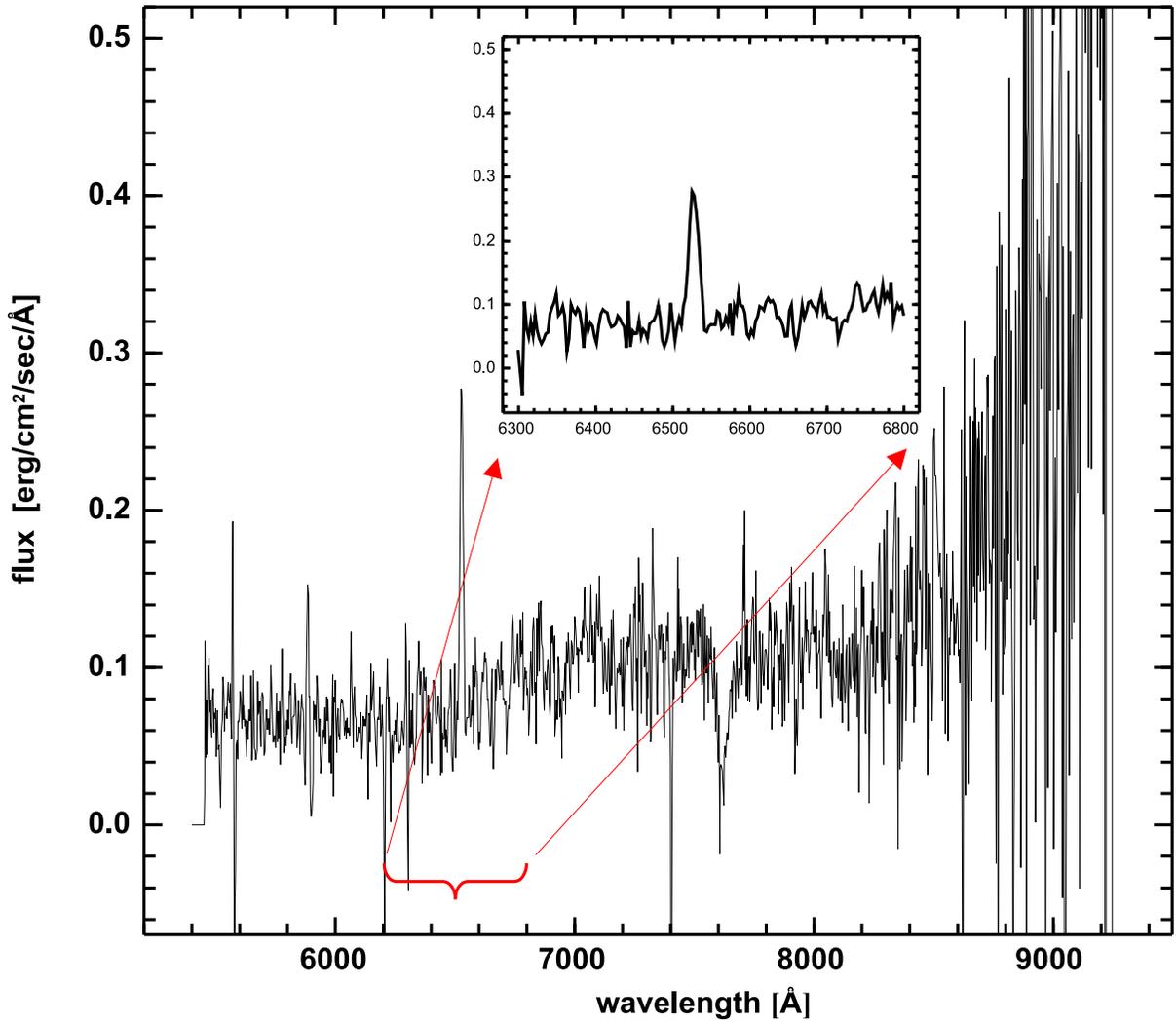}
\figcaption{
Spectrum of galaxy G north of quasar Q1253+104 showing a single prominent
emission line at 6527\,\AA. If this line is due to [O\textsc{ii}]\,3727 the
redshift of the galaxy is 0.75.\label{fig:1253_spec}}
\end{figure}

Spectra of the galaxy labled G in \fig \ref{fig:1253_opt} were taken with 
MOSCA on 2000 February
9 and 10. In that run MOSCA was equipped with a SITe CCD of
$2\mathrm{k}\times4\mathrm{k}$ $15\,\mu$ pixels. Grism green-500 was used together with an order
separation GG495 filter. The dispersion was 2.75\,\AA\ per pixel covering the
wavelength range 5500 to 9200\,\AA. For a rough flux calibration the standard
star HZ\,2 was observed. Three separate spectra with 3000\,sec integration time
were taken. The analysis of the spectra was carried out with context
\texttt{long} in MIDAS. Figure \ref{fig:1253_spec} shows the reduced and 
summed spectrum. A
prominent emission line is seen at a wavelength of 6527.0\,\AA. This line was
already detected with spectra taken with the same instrument in February 1997
and June 1998, although the quality of those spectra was inferior to the ones
presented here.

The line cannot be due to [O\textsc{iii}]4959,5007 as we do not see the other
line of this duplet. If it is due to [O\textsc{ii}]\,3727 we should see the
[O\textsc{iii}] lines in the red wavelength range around 8700\,\AA. We have
taken two 3000\,sec exposures with the red-1000 grating covering the range from
5000 to above 9000\,\AA. Unfortunately this region is heavily contaminated with
night sky lines, so we cannot exclude that the line seen in the green-500
spectrum is indeed [O\textsc{ii}]\,3727. This would then argue for a redshift
of 0.75 for the galaxy. H$\beta$ at 8513\,\AA\ is, however, also not seen. No
significant lines were seen in the blue spectral range observed with the
blue-500 grating in two 3000\,sec exposures.

\section{MODELING OF 1253+104: THE NULL RESULT}

\begin{figure}
\plotone{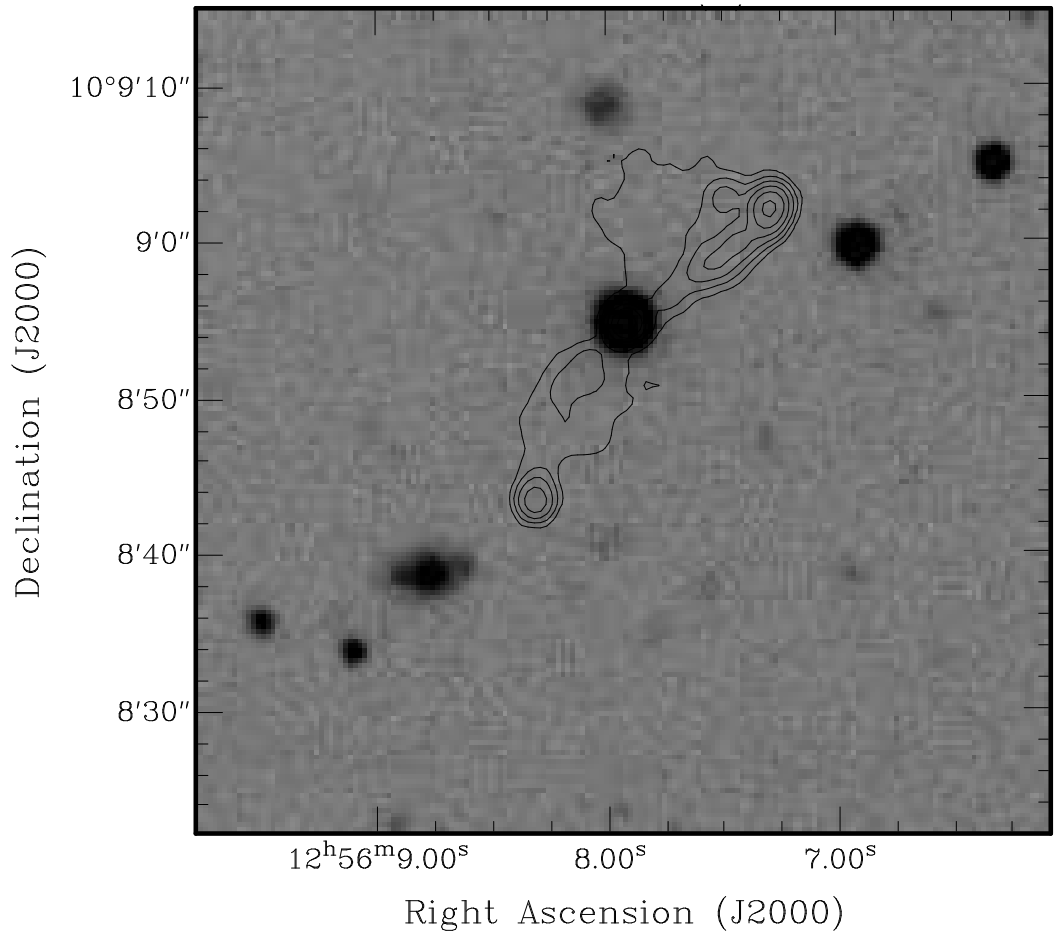}
\figcaption{\label{Fig:1253+104.opt.AB}The optical field of 1253+104 in R (left)
and B (right) bands, taken using the 3.5m telescope at the Calar Alto
observatory. Spectroscopy of the north-east galaxy places it at a
redshift of 0.75. The contours are of the VLA C-band
radio observations.}
\end{figure}

\fig \ref{Fig:1253+104.opt.AB} shows the optical field of 1253+104
with the radio emission drawn as contours.
The sources to the South-West are closely aligned
with the jet axis and would produce very little \( \eta _{G} \) signal.
The sources to the North-East and North-West, however, are off-axis
and close enough to be able to produce an \(\eta_{G}\) signal. However, the 
sources to the
east are most likely stellar and the extended source to the North-West
is likely at a redshift of 0.75 (see \S \ref{sect:Observations}), which is very 
close to the quasar's redshift.

\fig \ref{fig:3c9+1253} shows the North-East jet, its polarization
structure and the choice of the fiducial line which was computed
by averaging the intensity contours. One can clearly see that the
polarization follows the morphology quite closely. Using the same
methods described above, we compute \( \eta _{G} \),
which is plotted in \fig \ref{Fig:etag.3c9}. As one can see,
the \( \eta _{G} \) signal is very weak and, in fact, consistent
with zero to within the error bars. 

We conclude that there is no detectable gravitational lensing
in the jet of QSO1253+104. This is entirely consistent with the fact
that the only viable lens was found to be at a comparable redshift to that 
of the radio jet, so that $D_l \simeq 0$ (see equation (\ref{Eq:etag_theo})).
While this constitutes a null result, it is still an important null
result, in that it establishes that the morphology and
polarization do indeed follow one another. We can also be more confident
that our method of determining the uncertainty in \( \eta _{G} \)
is appropriate, since our initial measurement produced a signal that
was consistent with zero to within the uncertainties.

\section{COMMENTS}
The observed $\eta_G$ is very sensitive to the choice of representative
points along the jet.  
Furthermore, once the jet points are determined, they remain
fixed for all subsequent modeling, so it is worth investing some time in 
determining the ``best'' representative points.  Through simulations,
we find that using the
average behavior of the isophotal contours followed by a curvature
minimization algorithm consistently reproduces reasonable values of $\eta_G$.
This methodology requires very little human intervention and is therefore
more objective than choosing jet points by eye.

In the most favorable situation where all lenses are identified and
their positions in angular and redshift space are measured, $\eta_G$ can
yield good estimates of the mass of a galaxy that is interior to the largest 
impact
parameter $R_o$.  One important advantage of the $\eta_G$ technique is that
it measures a continuous range of impact parameters for the lens, ie., a jet can probe
both large and small impact parameters. Therefore, in principle one can also 
directly measure the
mass profile of individual galaxies.  We also find that the ellipticity of the
lens's mass distribution usually has very little effect on $\eta_G$ and can 
therefore be 
neglected in the determination of the lens parameters.

\section{CONCLUSIONS}

The $\eta_G$ technique is a useful astrophysical tool for measuring the
mass of galaxy-sized gravitational lenses.  With the more than 600 
quasars and AGN known to have radio jets \citep{Liu:2002}, we have an 
excellent laboratory
in which to measure the physical parameters of advantageously placed 
intervening galaxies.
The numbers of measurable, 
individual mass intervenors will
improve
with the next generation of radio and optical telescopes.  With a large
enough sample, one could in principle begin to constrain the relevant 
cosmological
parameters using statistical methods (see \citet{Falco:98,Keeton:2002}).

In the case of 3C9, we have used a more statistically robust method to
find the best fit lens parameters of two intervening galaxies and place
confidence intervals on their values.  For galaxy G1, we recover the results
of \citet{Kronberg:96}:  total mass of $17 \pm 10\ \mgal$ and an upper limit
of 75 for the mass-to-light ratio.  For galaxy G2, which has no measured
redshift, we can place an overall lower limit on the mass:  $10 \mgal$,
corresponding to a redshift of $z_2 = 0.25$.
However, this would lead to an unrealistic mass-to-light ratio and we
conclude that it is more likely the galaxy has a redshift $z_2\ga 1$.
If one identifies this galaxy with the absorption feature at $z_{abs}=1.6$,
this leads to $M_{G2} = 65\pm 15\ \mgal$ and $\ML \la 30$.  Also, due to the
observed lack of multiple images, we can place a lower limit on the core
mass of G2:  $a_{G2} > 5\ \kpc$.

While it is important to find as many gravitational lenses as possible, 
establishing null results like QSO 1253+104 is also important.  In this way,
we can determine a statistical ``baseline'' of the intrinsic, non-gravitational
deviations 
between the morphology and polarization ($\kappa$)
and hence better determine the level of significance of any detections
of gravitational lensing.

The quasar 3C9 provides a good example of the methodology we have developed
and illustrates the strengths and limitations of the technique.  QSO 1253+104
provides a useful null result in that our methodology correctly measures an
$\eta_G$ consistent with no gravitational lensing.

\acknowledgements{
We especially thank the director and staff of the Calar Alto Observatory and
the Max-Planck-Institut f\"{u}r Astronomie, Heidelberg for their hospitality,
and for the generous allotment of time on the Calar Alto 3.5m. telescope.  CRB,
CCD and PPK wish to acknowledge the Natural Sciences and Engineering Research
Council of Canada for support through their post-graduate scholarship (CRB) and
research grant programs (CCD and PPK).  Finally, we wish to thank the anonymous
referee for constructive comments and Margaret Burbidge for assistance at 
early stages of this project.}

\clearpage

%

\clearpage
\begin{deluxetable}{lccccc}
\tablecolumns{6}
\tablecaption{Derived parameters for the 3C9 lenses. \label{tab:params}}
\tablehead{
   \colhead{Assumed \(z_2\)}& \colhead{\(M_{G1}\)} & 
   \colhead{\(\left(\ML\right)_{G1}\)} & \colhead{\(M_{G2}\) } &
   \colhead{\(\left(\ML\right)_{G2}\)} & \colhead{\(a_{G2}\)}\\
   \colhead{} & \colhead{\(10^{11}M_\odot\)} & \colhead{\(M_\odot/L_\odot\)} &
   \colhead{\(10^{11}M_\odot\)} & \colhead{\(M_\odot/L_\odot\)} & 
   \colhead{\kpc}} 
\startdata
\(1.0\) & \(17 \pm 15\) & \(<75\)\tablenotemark{a}& \(25 \pm 10 \) & \(\la 40\) & 
\(> 5\) \\
\(1.6\) & \(17 \pm 15\) &\(<75\)\tablenotemark{a} & \(65\pm 15\) & \( \la 30 \) &
\(> 5\)\\
\(0 < z_2 < 2 \)&\(17 \pm 10\) &\(<75\)\tablenotemark{a} & \(\ga 10 \) & \( \ga 20 \) &
\nodata \\
\enddata
\tablenotetext{a}{Taken from \citet{Kronberg:96}.}
\end{deluxetable}

\end{document}